\documentclass[aps,pre,reprint,superscriptaddress,nofootinbib]{revtex4-2}
\usepackage{amsmath,amssymb,graphicx}
\usepackage[colorlinks=true,linkcolor=blue,citecolor=blue,urlcolor=blue]{hyperref}
\newcommand{\Tr}{\mathrm{Tr}}
\begin{document}
\title{Entanglement asymmetry in the gapped XYZ spin-$\frac12$ chain}
\author{Felipe Taha Sant'Ana}
\email{t.felipe@ufabc.edu.br}
\affiliation{Center of Mathematics, Computing, and Cognition, Federal University of ABC, 09210-580, Santo Andr\'e, Brazil}
\date{\today}
\begin{abstract}
The entanglement asymmetry measures how strongly a symmetry is broken inside a subsystem. Analytic
results at equilibrium have so far covered free theories and, perturbatively, the critical XXZ chain.
We compute the R\'enyi entanglement asymmetries of a large interval in the gapped, $U(1)$-breaking
phase of the interacting XYZ chain. The calculation
combines three ingredients. A charged-moment identity, which we prove for fermionic Gaussian and for
injective matrix-product ground states, ties the asymmetry to the static susceptibility of the broken
charge. A non-conservation sum rule then evaluates the susceptibility from sine-Gordon form factors,
its two-kink and one-breather channels providing a lower bound on the universal amplitude. The Baxter--Johnson--Krinsky--McCoy solution supplies the kink mass for different couplings. Infinite-system density-matrix renormalization group simulations built on these masses reproduce the master formula.
\end{abstract}
\maketitle

\section{Introduction}
\label{sec:intro}
The entanglement asymmetry~\cite{AMC23} (see~\cite{AresRev} for a review) measures how far a subsystem
departs from a symmetry of the full state. It has already helped explain anomalous relaxation,
including the quantum Mpemba effect and its microscopic origin~\cite{RKACMB24}, observed in trapped-ions experiments~\cite{Joshi24}. At equilibrium, it diagnoses symmetry breaking directly.

We recall the definition of Ref.~\cite{AMC23}. Let $\rho_A$ be the reduced density matrix of a
subsystem $A$, and let $Q_A$ be the part of a $U(1)$ charge supported in $A$, with $\Pi_q$ the
projector on its eigenspace of charge $q$. Deleting from $\rho_A$ every coherence between different
charge sectors gives the symmetrized state
\begin{equation}
\rho_{A,Q}=\sum_q \Pi_q\rho_A\Pi_q,
\label{eq.rhoAQ}
\end{equation}
which commutes with $Q_A$ and has the same charge distribution as $\rho_A$. The R\'enyi entanglement
asymmetry is the difference between the R\'enyi entropies of the two states~\cite{AMC23,MAKC23},
\begin{equation}
\begin{aligned}
\Delta S^{(n)}_A&=S^{(n)}(\rho_{A,Q})-S^{(n)}(\rho_A)\\
&=\frac{1}{1-n}\log\frac{\mathrm{Tr}\rho_{A,Q}^n}{\mathrm{Tr}\rho_A^n},
\end{aligned}
\label{eq.def}
\end{equation}
where
\begin{equation}
S^{(n)}(\rho)=\frac{1}{1-n}\log\mathrm{Tr}\rho^n,\qquad n>0,\quad n\neq1,
\label{eq.renyi}
\end{equation}
is the R\'enyi entropy of index $n$. As $n\to1$ it tends to the von Neumann entropy
$S(\rho)=-\mathrm{Tr}\rho\log\rho$, and $\Delta S_A\equiv\lim_{n\to1}\Delta S^{(n)}_A$ becomes the relative
entropy $\mathrm{Tr}[\rho_A(\log\rho_A-\log\rho_{A,Q})]$ between the state and its symmetrization. The
block projection~\eqref{eq.rhoAQ} can only mix the spectrum of $\rho_A$ (the spectrum of $\rho_{A,Q}$ is
majorized by that of $\rho_A$), so $\Delta S^{(n)}_A\ge0$ for every $n$, with equality if and only if
$[\rho_A,Q_A]=0$~\cite{AMC23}. These two properties make $\Delta S^{(n)}_A$ a measure of symmetry breaking
inside $A$. For integer $n\ge2$, $\mathrm{Tr}\rho_{A,Q}^n$ is a trace over $n$
copies of the subsystem. This replica structure is how we compute the asymmetry in
Sec.~\ref{sec:master}, and the $n=2$ asymmetry is the quantity accessed in the trapped-ion experiment
of Ref.~\cite{Joshi24}. The von Neumann value follows by continuation in $n$.

When the ground state breaks a $U(1)$ symmetry explicitly, $\Delta S^{(n)}_A$ grows as $\frac12\log\ell$
with the length $\ell$ of the subsystem, and the additive constant encodes the breaking. This constant
is known analytically only for free fermions~\cite{MAKC23,FAC23} and, from conformal perturbation theory
near the isotropic point, for the critical XXZ chain~\cite{Lastres24,Fossati24}. In the gapped case it
has been reached only for discrete groups, and only in the free Ising field theory~\cite{CM23}, through
the composite twist-field bootstrap~\cite{HCCA22}. Kink asymmetry on the free Ising chain was treated
in~\cite{Khor24}. We take up an interacting example: the XYZ chain in its $\gamma$-driven gapped phase,
where the anisotropy breaks $U(1)_z$.

We obtain the constant as a two-kink$+B_1$ lower bound ($B_1$ the first breather) plus a multi-particle
tail that two moment sum rules confine to a two-sided corridor on the attractive side of the
free-fermion point ($\beta^2 = 1/2$) and bound from below on the repulsive side. At $\beta^2 = 1/2$, the
amplitude follows a tangent law. The leading piece of that tail, the four-kink term, vanishes at the
free-fermion point; it is bounded from below and estimated perturbatively, and accounts for the small
excess seen in infinite-system density-matrix renormalization group (iDMRG) simulations. We also
confirm the master formula itself directly.

The argument has three parts, and only the last two are specific to the XYZ chain.
Section~\ref{sec:master} ties the asymmetry of a large interval to the static susceptibility
$\chi_{zz}$ of the broken charge. This step uses the Gaussian statistics of the subsystem charge and the
exponential clustering of the ground state; we prove it for fermionic Gaussian states and for injective
matrix-product states, and Sec.~\ref{sec:scope} states which of its ingredients reach beyond the XYZ
chain. Section~\ref{sec:sumrule} evaluates $\chi_{zz}$ in the sine-Gordon scaling limit through a
spectral sum rule, and Sec.~\ref{sec:lattice} fixes the lattice mass scale from the
Baxter--Johnson--Krinsky--McCoy solution. Section~\ref{sec:numerics} compares each step with iDMRG.
Longer derivations are collected in the appendices.

\section{Asymmetry and the fluctuations of the broken charge}
\label{sec:master}
\subsection{Model}
We consider the system described by the Hamiltonian
\begin{equation}
H=\sum_j\left[(1{+}\gamma)\sigma^x_j\sigma^x_{j+1}+(1{-}\gamma)\sigma^y_j\sigma^y_{j+1}
+\Delta_z\sigma^z_j\sigma^z_{j+1}\right],
\label{eq.H}
\end{equation}
with $|\Delta_z|<1$ and $0<\gamma$, in the thermodynamic-limit ground state (one N\'eel-$x$ vacuum). At
$\gamma=0$ this is the XXZ chain, which conserves $Q=\frac12\sum_j\sigma^z_j$. The anisotropy term
$\gamma(\sigma^x_j\sigma^x_{j+1}-\sigma^y_j\sigma^y_{j+1})=2\gamma(\sigma^+_j\sigma^+_{j+1}+\sigma^-_j\sigma^-_{j+1})$
changes $Q$ by $\pm2$ and breaks $U(1)_z$ explicitly. The broken charge on an interval $A$ of $\ell$
sites is $Q_A=\frac12\sum_{j\in A}\sigma^z_j$; we take $\ell$ even, so that $Q_A$ has integer
eigenvalues.

\subsection{Charged moments and the Gaussian saddle}
\label{sec:saddle}
Writing each projector in Eq.~\eqref{eq.rhoAQ} through its Fourier representation,
$\Pi_q=\int_{-\pi}^{\pi}\frac{d\alpha}{2\pi}e^{i\alpha(Q_A-q)}$, turns the moments of $\rho_{A,Q}$ into
circle averages of charged moments~\cite{AMC23,MAKC23}. For $n=2$,
\begin{equation}
\begin{aligned}
\frac{\mathrm{Tr}\rho_{A,Q}^2}{\mathrm{Tr}\rho_A^2}&=\int_{-\pi}^{\pi}\frac{d\alpha}{2\pi}Z_2(\alpha),\\
Z_2(\alpha)&=\frac{\mathrm{Tr}[\rho_A e^{i\alpha Q_A}\rho_A e^{-i\alpha Q_A}]}{\mathrm{Tr}\rho_A^2}.
\end{aligned}
\label{eq.Z2circle}
\end{equation}
Expanding $Z_2$ to $\mathcal O(\alpha^2)$ in the eigenbasis $\rho_A|n\rangle=p_n|n\rangle$ gives
$Z_2(\alpha)=1-V\alpha^2+\mathcal O(\alpha^4)$ with the width
\begin{equation}
V=\frac{1}{2\mathrm{Tr}\rho_A^2}\sum_{n\neq m}|\langle n|Q_A|m\rangle|^2(p_n-p_m)^2
\label{eq.V}
\end{equation}
[Appendix~\ref{app.sec.charged_moments}]. Each off-diagonal matrix element of $Q_A$ enters with the
squared difference of the two probabilities it connects, so $V$ vanishes if and only if
$[Q_A,\rho_A]=0$ and measures the coherence between charge sectors. When $\log Z_2$ is extensive in
$\ell$, the relevant phases are $\alpha\sim V^{-1/2}\sim\ell^{-1/2}$, the quartic term is suppressed by
$1/\ell$, and the circle average is a Gaussian integral,
\begin{equation}
\Delta S^{(2)}_A=-\log\int\frac{d\alpha}{2\pi}e^{-V\alpha^2}=\frac12\log(4\pi V).
\label{eq.saddle}
\end{equation}
The neighbourhood of $\alpha=\pm\pi$, where $e^{i\pi Q_A}$ is a $\sigma^z$ string, contributes terms
exponentially small in $\ell$ (Sec.~\ref{sec:num_direct}).

\subsection{From the width to the charge variance}
\label{sec:VVar}
The width $V$ is not the variance of $Q_A$ in $\rho_A$. In the same eigenbasis,
$\mathrm{Var}(Q_A)=\mathrm{Tr}[\rho_AQ_A^2]-(\mathrm{Tr}[\rho_AQ_A])^2$ contains the same off-diagonal
matrix elements weighted by $p_n$, plus the spread of the diagonal ones. The relation
\begin{equation}
V=\mathrm{Var}(Q_A)[1+\mathcal O(\ell^{-1})]
\label{eq.VVar}
\end{equation}
reduces the asymmetry to a ground-state correlation function. We establish it for two classes of states.

For a fermionic Gaussian $\rho_A$ and a one-body $Q_A$, Wick's theorem writes both $V$ and
$\mathrm{Var}(Q_A)$ as sums over pairs of entanglement modes with occupations $f_k$
[Appendix~\ref{app.sec.charged_moments}]. The two mode weights coincide whenever both occupations are
sharp, $f_k\in\{0,1\}$, and differ only for the soft modes, whose single-particle entanglement energies
are $\mathcal O(1)$. In a gapped chain the single-particle entanglement spectrum has $\mathcal O(1)$ level
spacing, so the number of soft modes stays bounded as $\ell$ grows. Each contributes $\mathcal O(1)$ to
$V-\mathrm{Var}(Q_A)$, against an extensive common part.

For an injective matrix-product state (MPS) with right-canonical tensors $B$ and a gapped transfer
matrix, the variational class of our simulations, both quantities are governed by the phase-dressed
doubled transfer operator of a unit cell,
\begin{equation}
\mathcal T_\alpha=\sum_{\vec s}e^{-i\alpha m_{\vec s}}B^{\vec s}\otimes\bar B^{\vec s},
\label{eq.Talpha}
\end{equation}
with $m_{\vec s}$ the charge of the cell configuration. Its leading eigenvalue $\lambda(\alpha)$ is
nondegenerate and analytic near $\alpha=0$, with $\lambda(0)=1$. The generating function of the
subsystem charge is a boundary contraction of $\mathcal T_{-\alpha}^{L}$ over the $L$ cells of $A$, so
$\log\langle e^{i\alpha Q_A}\rangle=L\log\lambda(-\alpha)+\mathcal O(1)$. The charged moment contains the
same bulk contraction once for each copy of $\rho_A$, and $\log Z_2(\alpha)=2L\log|\lambda(\alpha)|+\mathcal O(1)$.
Both second derivatives at $\alpha=0$ are therefore fixed by the single curvature
$-\partial_\alpha^2\log|\lambda(\alpha)|$, which gives Eq.~\eqref{eq.VVar}
[Appendix~\ref{app.sec.charged_moments}]. This is a theorem for injective MPS of finite bond dimension.
Its application to the XYZ ground state rests on the fact that a gapped, exponentially clustering
one-dimensional ground state is approximated by this class with an error that decreases rapidly in the
bond dimension.

\subsection{General R\'enyi index}
\label{sec:renyi_n}
For integer $n$, inserting the Fourier representation for each of the $n$ projectors gives
\begin{equation}
\frac{\mathrm{Tr}\rho_{A,Q}^n}{\mathrm{Tr}\rho_A^n}=\int\prod_{j=1}^{n}\frac{d\alpha_j}{2\pi}Z_n(\vec u),\qquad u_j=\alpha_j-\alpha_{j+1},
\label{eq.Zn}
\end{equation}
where $\alpha_{n+1}\equiv\alpha_1$, so that $\sum_ju_j=0$, and $Z_n(\vec u)$ is the trace of
$\rho_Ae^{iu_1Q_A}\cdots\rho_Ae^{iu_nQ_A}$ divided by $\mathrm{Tr}\rho_A^n$. To second order, $\log Z_n$ couples the phases
on replicas $r$ and $r'$ through widths $V_d$ that may depend on their separation $d$ on the replica
ring [Eq.~\eqref{app.eq.Vd}], with $V_d=V_{n-d}$; at $n=2$ the single width is $V$. If all $V_d$ equal
$V$, the exponent is $-\frac V2\sum_j(\alpha_j-\alpha_{j+1})^2$, the quadratic form of the Laplacian of
the cycle graph on the $n$ replicas, and the Gaussian integral gives
\begin{equation}
\begin{aligned}
\frac{\mathrm{Tr}\rho_{A,Q}^n}{\mathrm{Tr}\rho_A^n}&=\frac{1}{\sqrt n}(2\pi V)^{-(n-1)/2},\\
\Delta S^{(n)}_A&=\frac12\log(2\pi V)+\frac{\log n}{2(n-1)}.
\end{aligned}
\label{eq.renyiGauss}
\end{equation}
The $n$-dependent constant is geometric. The Laplacian has a zero mode along
$\alpha_1=\dots=\alpha_n$, which contributes the length $2\pi\sqrt n$ of the torus in that direction, and
$n-1$ nonzero eigenvalues $4\sin^2(\pi k/n)$ whose product is $n^2$; together they give the factor
$n^{-1/2}$.

For $n\le3$ the reflection symmetry $V_d=V_{n-d}$ already makes all widths equal. For general $n$,
uniformity follows from the transfer operator. In $Z_n(\vec u)$ the phase $e^{iu_rQ_A}$ dresses only the
$r$-th sheet, and in the bulk of the interval the $n$ sheets are contracted by
$\bigotimes_r\mathcal T_{u_r}^{L}$. The cyclic gluing of the sheets, the only place where they couple,
is absorbed into boundary tensors at the two ends of $A$. The extensive part of $\log Z_n$ is then
$L\sum_r\log\lambda(u_r)$ [Eq.~\eqref{app.eq.lnZn}], a sum of functions of the individual $u_r$. Its
mixed derivatives $\partial_{u_r}\partial_{u_{r'}}$ with $r\neq r'$ vanish, and its linear term drops out
by $\sum_ru_r=0$. The extensive quadratic form is therefore $-\frac12\mathrm{Var}(Q_A)\sum_ru_r^2$, the
cycle-graph Laplacian with a single coefficient, while all dependence on $d$ sits in a boundary Hessian
that stays finite as $\ell\to\infty$. Hence $V_d=\mathrm{Var}(Q_A)[1+\mathcal O(1/\ell)]$ for every $d$
and every integer $n\ge2$ [Appendix~\ref{app.sec.general_renyi}]. At the free-fermion point, where the
widths can be computed from the Gaussian reduced density matrix, the differences $V_d-V_1$ saturate to
$\ell$-independent constants, as the proof requires; density matrices constructed explicitly from the
infinite MPS (iMPS) confirm this saturation [Appendix~\ref{app.sec.general_renyi}].

\subsection{Master formula}
Combining Eqs.~\eqref{eq.VVar} and \eqref{eq.renyiGauss} with $\mathrm{Var}(Q_A)\simeq\ell\chi_{zz}/2$,
at general R\'enyi index,
\begin{equation}
\Delta S^{(n)}_A(\ell)=\frac12\log(\pi\ell\chi_{zz})
+\frac{\log n}{2(n-1)}+\mathcal O\left(\frac{1}{M\ell}\right),
\label{eq.master}
\end{equation}
where $\chi_{zz}=2\lim_{\ell\to\infty}\mathrm{Var}(Q_A)/\ell=\frac12\sum_r\langle
\sigma^z_0\sigma^z_r\rangle_c$ and $M$ is the kink mass (the inverse kink length in lattice units);
the $\sigma^z\sigma^z$ correlation length is $1/2M$ on the repulsive side and $1/m_{B_1}$ on the
attractive side [Eq.~\eqref{eq.masses}]. The von Neumann limit $n\to1$ follows by analytic continuation
of the integer-$n$ result, $\Delta S_A=\frac12\log(\pi e\ell\chi_{zz})$, and is compared with the
numerics in Fig.~\ref{fig.asym}.

\subsection{Scope of the master formula}
\label{sec:scope}
Equation~\eqref{eq.master} combines a model-independent statement with one model-specific number. The
derivation up to Eq.~\eqref{eq.master}, with $\chi_{zz}$ read as the density of charge fluctuations, uses
three properties of the state: $Q_A$ is a sum of on-site charges with integer spectrum; $\log Z_n$ is
extensive with a quadratic leading term, so that the subsystem charge has Gaussian statistics on the
scale $\sqrt\ell$; and the ground state clusters exponentially, which bounds the number of soft
entanglement modes in the Gaussian proof and gaps the transfer matrix in the MPS proof. None of these
steps refers to the XYZ chain. The large-interval form $\frac12\log(\pi\ell\chi)+\log n/[2(n-1)]$, with
$\chi$ the susceptibility of the broken charge, is therefore expected for the ground state of any gapped
one-dimensional chain in which a $U(1)$ charge is broken explicitly, to the same extent as for the XYZ
ground state: the Gaussian proof covers free-fermion chains, and the MPS proof covers every state in the
injective class. In particular, the replica dependence carries no information on
the model, and the difference $\Delta S^{(2)}_A-\Delta S^{(3)}_A\to\frac12\log2-\frac14\log3$ is invariant.

What is specific to the XYZ chain is the value of $\chi_{zz}$, and its evaluation in
Sec.~\ref{sec:sumrule} has two layers of its own. The spectral representation of $\chi_{zz}$ and the
Cauchy--Schwarz inequalities that turn two further spectral moments into a corridor use only the
positivity of the spectral weight; they apply to any gapped theory in which the broken charge has a
local time derivative. The closed forms of those two moments use the bosonized form of the charge and
the sine-Gordon equation of motion, and the two-kink and breather terms use the sine-Gordon form
factors of the $q=2$ vertex through which $\gamma$ enters. The lattice mass scale of
Sec.~\ref{sec:lattice} comes from the integrability of the XYZ chain.

\section{The susceptibility from the sine-Gordon sum rule}
\label{sec:sumrule}
\subsection{Spectral representation}
\label{sec:spectral}
The scaling limit of the $\gamma$-perturbed XXZ chain is the sine-Gordon model, which we write in
Lukyanov--Zamolodchikov (LZ) conventions~\cite{LZ97},
$\mathcal A=\int d^2x\,[\frac1{16\pi}(\partial\phi)^2-2\mu\cos\beta\phi]$, with
$\beta^2=1-\theta_0/\pi$, $\theta_0=\arccos\Delta_z$, $\xi_{\rm sG}=\beta^2/(1-\beta^2)$. The spectrum
contains kinks of mass $M$ and, on the attractive side $\xi_{\rm sG}<1$ ($\Delta_z<0$), breathers; the
free-fermion point is $\xi_{\rm sG}=1$ ($\Delta_z=0$), and $\xi_{\rm sG}>1$ is the repulsive side. The
breaking vertex $e^{i\beta\phi}$ (the uniform part of $\sigma^+\sigma^+$) carries charge $q=2$. The
lattice generator maps to $Q=(q/\beta)\int dx\,\Pi$ with $\Pi=\partial_t\phi/4\pi$. This normalization
is fixed at the XXZ point, where $Q$ generates the symmetry; the $\gamma$-perturbation renormalizes only
the breaking vertex and leaves $q=2$ untouched [Appendix~\ref{app.sec:sumrule}].

Since the charge is broken,
\begin{equation}
\dot Q=i[H,Q]=-2q\mu\int dx\,\sin\beta\phi\neq0.
\label{eq.Qdot}
\end{equation}
For an eigenstate $|n\rangle$ of energy $E_n$ (ground state $E_0=0$),
$\langle n|\dot Q|0\rangle=iE_n\langle n|Q|0\rangle$, hence
$|\langle n|Q|0\rangle|^2=|\langle n|\dot Q|0\rangle|^2/E_n^2$. Inserting this into the completeness sum
$\mathrm{Var}(Q)=\sum_{n\neq0}|\langle n|Q|0\rangle|^2$ and using $\chi_{zz}=2\mathrm{Var}(Q)/L$ gives the
non-conservation sum rule
\begin{equation}
\chi_{zz}=\frac{2}{L}\sum_{n\neq0}\frac{|\langle n|\dot Q|0\rangle|^2}{E_n^2},
\label{app.eq.spectral}
\end{equation}
with the thermodynamic limit $L\to\infty$ taken first, at fixed intermediate-state content, and the
scaling limit afterwards. Every term is non-negative, so any subset of intermediate states bounds
$\chi_{zz}$ from below. The lightest states that couple to the $C$-odd source $\sin\beta\phi$ are the
two-kink (soliton--antisoliton) continuum above $2M$ and, for $\xi_{\rm sG}<1$, the first breather $B_1$
of mass $m_{B_1}=2M\sin(\pi\xi_{\rm sG}/2)$; the $C$-even states $B_2$ and $B_1B_1$ do not couple.

\subsection{Two-kink and breather channels}
\label{sec:channels}
With the universal ratio $\mathfrak c(\beta^2)\equiv\chi_{zz}/M$, the two-kink and one-breather
contributions are
\begin{align}
\mathfrak c_{2\rm k}&=\frac{q^2}{16\pi\beta^4}\int_{-\infty}^{\infty} d\theta\,
\frac{|\mathcal F(2\theta)|^2}{\cosh\theta},\qquad
\mathcal F(\theta)=\frac{\sinh\frac\theta2e^{\mathcal J(\theta)}}
{\cosh\frac{\theta+i\pi}{2\xi_{\rm sG}}},
\label{eq.c2k}\\
\mathfrak c_{B_1}&=\frac{(1+\xi_{\rm sG})^{2}}{2}
\left(\tan\frac{\pi\xi_{\rm sG}}{2}\right)^{-\xi_{\rm sG}}
e^{\eta(\xi_{\rm sG})}
\qquad(\xi_{\rm sG}<1),
\label{eq.cB1}
\end{align}
with
\[
\begin{aligned}
\mathcal J(\theta)&=-\int_0^\infty\frac{dt}{t}\frac{k_0(t)}{\sinh t}
\sin^2\frac{t(i\pi+\theta)}{2\pi},\\
k_0(t)&=\frac{\sinh[(\xi_{\rm sG}-1)t/2]}{\sinh(\xi_{\rm sG}t/2)\cosh(t/2)},
\end{aligned}
\]
and $\eta(\xi)=[\mathrm{Cl}_2(2\pi\xi)-4\mathrm{Cl}_2(\pi\xi)]/2\pi$, where
$\mathrm{Cl}_2(x)=\sum_{k\ge1}\sin(kx)/k^2$ is the Clausen function. For a two-kink state the
center-of-mass momentum delta in the form factor cancels the factor $1/L$ in
Eq.~\eqref{app.eq.spectral} and sets the two rapidities opposite, which leaves the single integral of
Eq.~\eqref{eq.c2k}.

Equation~\eqref{eq.c2k} follows from Lukyanov's two-soliton form factors of exponential
fields~\cite{Luk97,FT11}, assembled in the $C$-odd channel. Their normalization contains the vacuum
expectation value $\mathcal G_\beta=\langle e^{i\beta\phi}\rangle$, which the equation of motion
$\sin\beta\phi=-\Box\phi/(16\pi\mu\beta)$ trades for the kink mass and the coupling,
$\mathcal G_\beta\cot(\pi\xi_{\rm sG}/2)=M^2\xi_{\rm sG}/8\mu\beta^2$ [Eq.~\eqref{app.eq.EOM}]. The same
identity follows from the Hellmann--Feynman theorem and is equivalent to the known sine-Gordon bulk
energy $f=-\frac{M^2}{4}\tan(\pi\xi_{\rm sG}/2)$~\cite{Zam95,DdV95}. We also checked it against the LZ
vacuum expectation value~\cite{LZ97}, whose integral representation converges only on the attractive
side; for $\xi_{\rm sG}\ge1$ the identity supplies the continuation. Equation~\eqref{eq.cB1} comes from
the one-breather matrix element~\cite{Luk97} after the same elimination of $\mathcal G_\beta$, the
Clausen functions arising from $\int_0^{\pi\xi}t\,dt/\sin t$ [Appendix~\ref{app.sec:sumrule}]. The
breather carries the whole amplitude as $\xi_{\rm sG}\to0$, where $\mathfrak c_{B_1}\to\frac12$, and its
weight vanishes linearly at the free-fermion point, $\mathfrak c_{B_1}\simeq\pi(1-\xi_{\rm sG})$. At the
free-fermion point the two-soliton factor reduces to unity and the two-kink integral evaluates to one,
reproducing the known XY-chain result~\cite{MAKC23}; this fixes the constant prefactor common to both
channels.

Because the sum rule is a sum of squares and $\mathcal F$ is the minimal solution of the form-factor
equations, the combination $\mathfrak c_{2\rm k}+\mathfrak c_{B_1}$ is a lower bound on $\mathfrak c$.
The minimality of $\mathcal F$ is needed here: a nontrivial Castillejo--Dalitz--Dyson (CDD) factor would
either violate the ultraviolet growth bound, whose margin $(\xi_{\rm sG}-1)^2/4\xi_{\rm sG}(\xi_{\rm sG}+1)$
closes only at the free-fermion point, or introduce poles absent from the two-kink kinematics
[Appendix~\ref{app.sec:sumrule}]. The deficit of the bound is the weight of states with three or more
particles: $s\bar sB_1$ and heavier states on the attractive side, four kinks and heavier states on the
repulsive side.

\subsection{Moment sum rules and the tail corridor}
\label{sec:corridor}
Two further moments of the same spectral weight are known in closed form, and they make the deficit
two-sided [Appendix~\ref{app.sec:fourkink}]. We write
$T_j=\frac2L\sum_{n\neq0}E_n^{j}|\langle n|Q|0\rangle|^2$, so that $T_0=\chi_{zz}$.

The moment $T_{-1}$ is the static response to a uniform field coupled to the broken charge,
$H(h)=H-hQ$. Because $Q$ is built from the field momentum alone, completing the square and conjugating
with $e^{i\lambda\int dx\,\phi}$, which shifts $\Pi$ by a constant and commutes with every
$\phi$-dependent term of $H$, removes the field. The ground-state energy is then quadratic in $h$ to all
orders, and second-order perturbation theory gives
$T_{-1}=q^2/8\pi\beta^2=(1+\xi_{\rm sG})/2\pi\xi_{\rm sG}$ at all couplings [Eq.~\eqref{app.eq.Tm1exact}].
On the lattice this states that the uniform-field susceptibility of the gapped chain tends, as
$\gamma\to0$, to the XXZ value $K/\pi v$ with $K=1/2\beta^2$: the gap does not suppress the response,
because the charge is not conserved.

The moment $T_{+1}$ is the $f$-sum rule of the broken charge. The double commutator gives
$\frac12\langle[Q,[H,Q]]\rangle=q^2\mu L\mathcal G_\beta$, and the equation-of-motion identity turns it
into $T_{+1}=\frac14 q^2M^2(1+\xi_{\rm sG})\tan(\pi\xi_{\rm sG}/2)$ for $\xi_{\rm sG}<1$
[Eq.~\eqref{app.eq.Tp1exact}]. For $\xi_{\rm sG}\ge1$ the spectral sum diverges in the ultraviolet, in
step with the pole of $\tan(\pi\xi_{\rm sG}/2)$, and only $T_{-1}$ is available on the repulsive side.

Subtracting the two-kink$+B_1$ components from the three moments leaves the moments of a positive
measure $d\sigma(E)$ for the $\ge3$-particle tail, supported above the lightest omitted threshold
$E_{\rm th}$, which is $2M+m_{B_1}$ on the attractive side and $4M$ on the repulsive side:
$r_{-1}=\int d\sigma/E$, $M\mathfrak c_{\rm tail}=\int d\sigma$ and $M^2r_{+1}=\int E\,d\sigma$, with
$r_{\pm1}\ge0$. The pointwise bound $E\ge E_{\rm th}$ gives the lower edge, and the Cauchy--Schwarz
inequality $(\int d\sigma)^2\le\int d\sigma/E\int E\,d\sigma$ gives the upper one; with $E_{\rm th}$ in
units of $M$,
\begin{equation}
\begin{aligned}
E_{\rm th}r_{-1}&\le \mathfrak c_{\rm tail}\le\sqrt{r_{+1}r_{-1}} &&(\xi_{\rm sG}<1),\\
\mathfrak c_{\rm tail}&\ge 4r_{-1} &&(\xi_{\rm sG}>1).
\end{aligned}
\label{eq.corridor}
\end{equation}
Both edges are attained by admissible measures, all the weight at the single energy
$E^*=M\sqrt{r_{+1}/r_{-1}}$ for the upper one and weight at threshold plus weight escaping to infinity for
the lower one, so the two moments hold no further information on the tail. Numerically, $r_{-1}>0$ at
every coupling studied.

\subsection{The free-fermion point}
\label{sec:ffp}
At the free-fermion point the amplitude obeys the tangent law
\begin{equation}
\mathfrak c(\beta^2)=1+\left(\frac{\pi}{2}-1\right)(\xi_{\rm sG}-1)+ \mathcal O\left((\xi_{\rm sG}-1)^2\right)
\label{eq.ffp}
\end{equation}
from both sides. The two-kink term alone is not differentiable there: its one-sided slopes are
$\mathfrak c_{2\rm k}'(1^+)=\frac\pi2-1$ and $\mathfrak c_{2\rm k}'(1^-)=\frac{3\pi}2-1$. The difference
comes from the $B_1$ pole of $1/\cosh\frac{\theta+i\pi}{2\xi_{\rm sG}}$, which for $\xi_{\rm sG}$ near
one lies next to the two-kink threshold and carves out of $|\mathcal F|^2$ a notch of width
$\pi|\xi_{\rm sG}-1|/2$. The weight removed is linear in $|\xi_{\rm sG}-1|$ on both sides, so it enters
the two one-sided slopes with opposite signs. On the attractive side the slope $-\pi$ of
$\mathfrak c_{B_1}$ cancels the excess slope $\pi$ of $\mathfrak c_{2\rm k}$: the breather weight flows
continuously into the two-kink threshold as $\xi_{\rm sG}\to1^-$, which also tests Lukyanov's $B_1$
normalization [Appendix~\ref{app.sec:fourkink}].

At the free-fermion point, the source $\sin\beta\phi$ degenerates to the Dirac mass
$\bar\psi i\gamma^5\psi$. A free-field bilinear has form factors only in the
one-particle--one-antiparticle sector, so its connected four-particle form factor vanishes and every
$\ge3$-particle channel switches off. Away from that point the connected four-particle form factor is
generated at first order in the Thirring coupling $g=-\frac\pi2(\xi_{\rm sG}-1)+\dots$, and the tail,
quadratic in the form factor, switches on quadratically in the distance from the free-fermion point,
$\mathfrak c_{4\rm k}=A(\xi_{\rm sG}-1)^2+\dots$. A leading-order evaluation of the four-kink form
factor, the single-vertex Thirring amplitude integrated over the four-kink phase space, yields
$A\simeq0.08$ as a Monte-Carlo estimate [Appendix~\ref{app.sec:fourkink}]. Independently of that
integration, the tail moment vanishes as $r_{-1}=A_{-1}(\xi_{\rm sG}-1)^2$ with $A_{-1}\approx0.0030$
from both sides, and the repulsive edge of Eq.~\eqref{eq.corridor} bounds the coefficient from below,
$A\ge4A_{-1} \approx 0.012$. As the leading term in $\xi_{\rm sG}-1$, this estimate captures the small
excess seen near the free-fermion point.

\section{Exact lattice input: masses and coupling}
\label{sec:lattice}
The kink mass and the lattice-to-continuum coupling come from the Baxter/Johnson--Krinsky--McCoy (JKM)
solution~\cite{JKM73,Baxter,EEFR12,EEFR11,EER10}. Mapping $H$ onto the principal regime of the
eight-vertex model fixes an elliptic parametrization [Appendix~\ref{app.sec:deriv}], in which the kink
and lattice $B_1$ masses are
\begin{equation}
M=\mathrm{atanh}\,k_1',\qquad
m_{B_1}=2\left[\mathrm{atanh}\,k_1'+\log \mathrm{dn}(w;k_2')\right],
\label{eq.masses}
\end{equation}
with $k_2=\frac{1-k_1'}{1+k_1'}$, $w=\frac{2K(k_2)K(l')(\mu_{\rm e}-\pi/2)}{\pi K(l)}$, and the
modulus $l$ and Poincar\'e parameter $\mu_{\rm e}$ of Appendix~\ref{app.sec:deriv}.
On the XY line the Boltzmann anisotropy vanishes, a Landen transformation reduces the JKM modulus to
$k_1'=\gamma$, and Eq.~\eqref{eq.masses} gives $M=\mathrm{atanh}\,\gamma$. As $\gamma\to0$ the modulus
$k_2'$ vanishes, and the lattice breather relation tends to the sine-Gordon one,
$m_{B_1}/2M\to\sin(\pi\xi_{\rm sG}/2)$. In the same limit $K(l)$ diverges logarithmically in $\gamma$
while $K(l')\to\pi/2$, and a modular transformation to the complementary nome turns the elliptic
expressions into power laws,
\begin{equation}
M\to\left(\frac{\gamma\,4^{1-2\beta^2}}{\sin^2\theta_0}\right)^{\pi/2\theta_0},\quad
\mathcal C_{\sigma^+\sigma^+}(\Delta_z)=\kappa(\beta^2)\frac{4^{1-2\beta^2}}{1-\Delta_z^2},
\label{eq.amp}
\end{equation}
where $\mathcal C$ is the amplitude of the breaking coupling, $\mu=\gamma\mathcal C$, and
$\kappa(\beta^2)$ is the sine-Gordon mass--coupling constant~\cite{Zam95}. The exponent
$\pi/2\theta_0=1/(2-2\beta^2)$ is the sine-Gordon mass--coupling exponent, $M\propto\mu^{1/(2-2\beta^2)}$,
so matching the lattice mass to it fixes $\mathcal C$. We verify that Eq.~\eqref{eq.amp} matches the
limit at different couplings, with $\mathcal C(0)=1/2\pi$. Combining
Eqs.~\eqref{eq.master}--\eqref{eq.amp}, we obtain the leading $\gamma\to0$ entanglement asymmetry
(displayed across R\'enyi index and coupling in Fig.~\ref{fig.asym}):
\begin{equation}
\begin{aligned}
\Delta S^{(n)}_A(\ell)=&\frac12\log (\pi\ell)+\frac12
\log \left[\mathfrak c(\beta^2)\left(\frac{\gamma\,4^{1-2\beta^2}}{\sin^2\theta_0}\right)^{\pi/2\theta_0}\right] \\
&+\frac{\log n}{2(n-1)},
\end{aligned}
\label{eq.final}
\end{equation}
for $\gamma\to0$ and $\ell M \gg 1$.
The finite-$\gamma$ corrections to it are of two kinds. The first is
analytic, a regular power series in $\gamma$, the ordinary corrections to scaling. The second is
non-analytic and of umklapp type: lattice operators that are absent from the continuum
sine-Gordon description but are generated on the chain contribute a fractional power
$\gamma^{(4K-2)\pi/2\theta_0}$, whose exponent is fixed by the Luttinger parameter $K=1/2\beta^2$. Our
$\gamma\to0$ extrapolations include both. The kink mass, finally, need not be taken in the scaling
limit: Eq.~\eqref{eq.masses} supplies $M$ at any $\gamma$.

\section{Numerical validation}
\label{sec:numerics}
\subsection{Susceptibility and the amplitude}
\label{sec:num_c}
We ran a systematic iDMRG program~\cite{tenpy} in the symmetry-broken vacuum. The
susceptibility $\chi_{zz}(\gamma,\Delta_z)$ is measured from correlator sums that are stable in the
bond dimension, while the masses come from Eq.~\eqref{eq.masses}. We then extrapolate
$\mathfrak c(\gamma)=\chi_{zz}/M$ to $\gamma\to0$ with polynomial and umklapp-corrected forms, whose
spread sets the quoted errors. The fits are shown in Fig.~\ref{app.fig.extrap}. The extrapolation
reaches farther above the smallest-$\gamma$ data on the repulsive side than on the attractive side.

\begin{figure}[t]
\includegraphics[width=\columnwidth]{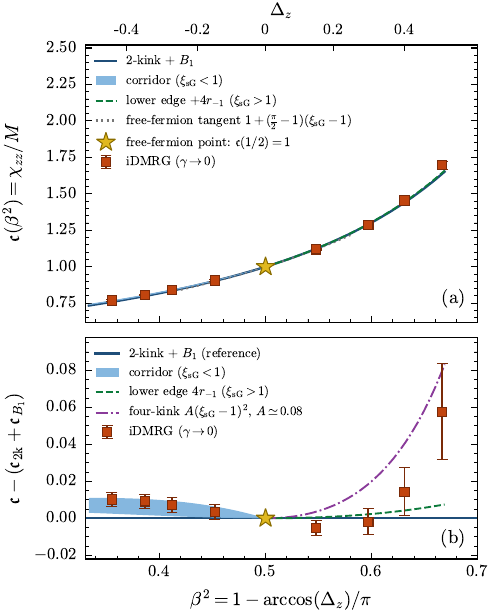}
\caption{The universal coefficient $\mathfrak c(\beta^2)=\chi_{zz}/M$. (a) Line: the two-kink$+B_1$
evaluation, Eqs.~\eqref{eq.c2k}--\eqref{eq.cB1}, a lower bound on $\mathfrak c$ (the $B_1$
component exists only for $\beta^2<\frac12$). Shaded band: the two-sided corridor,
Eq.~\eqref{eq.corridor}, for $\xi_{\rm sG}<1$ (sub-symbol on this scale); dashed: the corridor's lower
edge $\mathfrak c_{2\rm k}+4r_{-1}$ for $\xi_{\rm sG}>1$; dotted: the free-fermion tangent,
Eq.~\eqref{eq.ffp}. Squares: iDMRG determinations $\chi_{zz}/M_{\rm exact}$ at
eight couplings, using the exact mass, extrapolated to $\gamma\to0$ (Table~\ref{app.tab:grand}); star:
the free-fermion point $\mathfrak c(\frac12)=1$. (b) The same quantities measured from the
two-kink$+B_1$ curve, which becomes the horizontal line at zero: the tail corridor
$[E_{\rm th}r_{-1},\sqrt{r_{+1}r_{-1}}]$ for $\xi_{\rm sG}<1$ (band), its lower edge $4r_{-1}$ for
$\xi_{\rm sG}>1$ (dashed), the leading four-kink estimate $A(\xi_{\rm sG}-1)^2$ with $A\simeq0.08$
(dash-dotted, drawn for $\xi_{\rm sG}>1$), and the iDMRG excess
$\mathfrak c-(\mathfrak c_{2\rm k}+\mathfrak c_{B_1})$ with the error bars of panel (a).}
\label{fig.c}
\end{figure}

Table~\ref{app.tab:grand} and Fig.~\ref{fig.c} collect the outcome at eight couplings that sample the
broken phase from $\xi_{\rm sG}=0.55$ to $2$. On the scale of Fig.~\ref{fig.c}(a) the two-kink$+B_1$
curve carries nearly all of the coupling dependence, and the bounds, the tail estimate and the data lie
close to it. Figure~\ref{fig.c}(b) subtracts this curve, leaving the multi-particle tail and its bounds.
The two-kink$+B_1$ lower bound is satisfied to within the combined fit-and-extrapolation uncertainty at
every coupling, and the determinations fall inside the corridor of Eq.~\eqref{eq.corridor} at all four
attractive couplings [Table~\ref{app.tab:grand}]; e.g.\ $\mathfrak c\in[0.7613,0.7698]$ at
$\xi_{\rm sG}=0.55$ against the measured $0.769$. Two repulsive central values sit marginally below the
bound. Since the corridor's lower edge lies strictly above the bound there ($\mathfrak c\ge1.1247$ at
$\xi_{\rm sG}=1.212$, $\ge1.2931$ at $1.4813$), these negative excesses are identified as $\gamma\to0$
extrapolation systematics [Appendix~\ref{app.sec:fourkink}].

The excess over the bound is small. It is resolved at the largest repulsive coupling
$\xi_{\rm sG}=2$ and shows a small positive trend on the attractive side toward $\xi_{\rm sG}=0.55$,
while the intermediate repulsive points are consistent with the bound. It vanishes quadratically from
the free-fermion point, where the four-kink law $A(\xi_{\rm sG}-1)^2$ describes its onset and meets the
largest repulsive coupling. At the intermediate repulsive couplings the leading-order line overshoots
the measured excess, as expected for a leading term in $\xi_{\rm sG}-1$ whose finite-$\xi_{\rm sG}$
off-shell dressing is omitted. At $\xi_{\rm sG}=2$ the measured excess lies well above the lower edge
$4r_{-1}$, which requires most of the four-kink weight to sit far above threshold. The excess provides an
estimate of the multi-particle weight in the sum rule, which need not be symmetric about the
free-fermion point because the missing contributions differ on the two sides (four-kink versus
$B_1$-involving states).

\subsection{Direct evaluation of the asymmetry}
\label{sec:num_direct}
The asymmetry can also be computed directly from the iMPS. For right-canonical tensors and left Schmidt
weights $\lambda$, the charged moments of a bulk interval reduce to
$\sum\lambda_a^2\lambda_c^2|E^{(\ell)}_\alpha[(ac),(bd)]|^2$ with $E_\alpha$ the phase-dressed doubled
transfer matrix of Eq.~\eqref{eq.Talpha}; one pass per $\alpha$ yields all $\ell$ (see~\cite{CV23} for the
complementary pattern-function limit of MPS asymmetries). This construction reproduces the second
R\'enyi asymmetry $\Delta S^{(2)}_A(\ell)$ built directly from interval density matrices with
charge projectors. Specializing the master formula~\eqref{eq.master} to $n=2$ predicts a fixed
subleading constant, $\Delta S^{(2)}_A(\ell)=\frac12\log\ell+c_1+\mathcal O((M\ell)^{-1})$ with
$c_1=\frac12\log(2\pi\chi_{zz})$. Fitting the measured $\Delta S^{(2)}_A(\ell)$ across interval sizes at
the interacting point $(\Delta_z,\gamma)=(0.3,0.2)$ extracts $c_1\approx -0.1834$. Inverting this relation returns
$\chi_{zz}=e^{2c_1}/2\pi \approx 0.1103$, in agreement with the directly measured $0.1105$, corroborating
Eq.~\eqref{eq.master} in the interacting theory. The $\alpha=\pm\pi$ ($\sigma^z$-string)
region of the circle average is exponentially suppressed in the interval length
and provides no contribution.

Writing $c_1^{(n)}$ for the constant in $\Delta S^{(n)}_A(\ell)=\frac12\log\ell+c_1^{(n)}+\dots$, we
test the master formula along three different paths. At the free-fermion point $\Delta_z=0$, a
transfer-matrix evaluation of Eq.~\eqref{eq.def} extrapolates to the predicted $n=2$ and $n=3$
constants. In addition, at the interacting points, the asymmetry-route susceptibility matches the
directly measured variance, within the finite-window systematic quantified in
Appendix~\ref{app.sec:numerics}. And finally, the parameter-free difference
$c_1^{(2)}-c_1^{(3)}=\frac12\log 2-\frac14\log 3$, measured with $n=2$ and $n=3$ contractions on the same
state, is recovered.

\begin{figure*}[t]
\includegraphics[width=\textwidth]{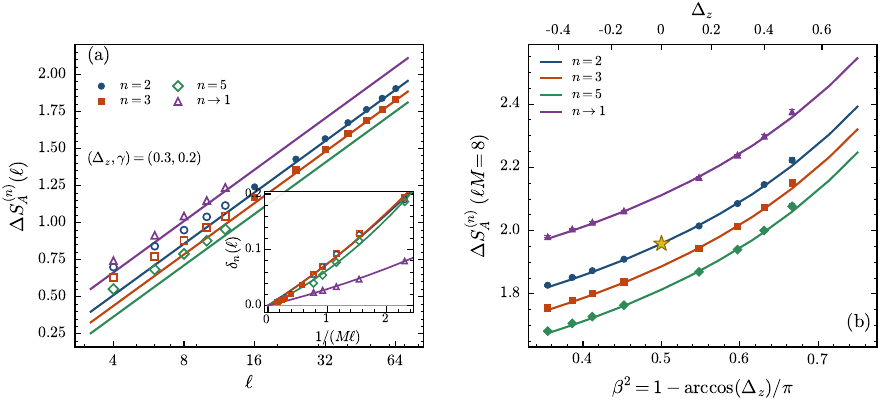}
\caption{The R\'enyi entanglement asymmetry $\Delta S^{(n)}_A$ itself, resolved in replica index and
coupling. (a) $\Delta S^{(n)}_A(\ell)$ versus interval length at the interacting point
$(\Delta_z,\gamma)=(0.3,0.2)$ for $n=2,3,5$ and the von Neumann limit $n\to1$. Symbols are iMPS data:
open, from interval reduced density matrices with charge projectors ($\ell\le12$); filled, from the
charged-moment construction ($\ell=16$--$64$, $n=2,3$). Lines correspond to the master formula
Eq.~\eqref{eq.master}. Inset: deviation $\delta_n(\ell)\equiv\Delta S_A^{(n)}(\ell)-[\frac12\log(\pi\ell\chi_{zz})+\log n/(2(n-1))]$ from the master formula, versus $1/(M\ell)$. Open (small $\ell$) and filled (large $\ell$) points fall on a common trend through the origin, indicating that the small-$\ell$ offset composes the $\mathcal{O}((M\ell)^{-1})$ tail with no constant remainder.
(b) $\Delta S^{(n)}_A$ at fixed $\ell M=8$ versus $\beta^2$ (bottom axis) and $\Delta_z$ (top axis), for
the same four indices. Lines: Eq.~\eqref{eq.final}, i.e., the master formula with
$\chi_{zz}=M\mathfrak c$, evaluated with the two-kink$+B_1$ coefficient of
Eqs.~\eqref{eq.c2k}--\eqref{eq.cB1} and without the $\mathcal O((M\ell)^{-1})$ term; the branches are
offset rigidly by $\log n/[2(n-1)]$. Symbols: the same expression evaluated with the $\gamma\to0$ iDMRG
determinations of $\mathfrak c$ (Table~\ref{app.tab:grand}), with error bars propagated from those of
$\mathfrak c$. Star: the free-fermion point, $\mathfrak c(\frac12)=1$, on the $n=2$ branch.}
\label{fig.asym}
\end{figure*}

We validate the master formula for each replica index; Fig.~\ref{fig.asym} displays the
asymmetry itself. Panel~(a) follows $\Delta S^{(n)}_A(\ell)$ at a representative interacting
point $(\Delta_z,\gamma)=(0.3,0.2)$ for $n=2,3,5$ and the von Neumann limit: projector density
matrices at small $\ell$ and the charged-moment construction at large $\ell$ collapse onto
Eq.~\eqref{eq.master}, sharing the $\frac12\log\ell$ growth and separated by the offsets
$\log n/[2(n-1)]$ that replica-width uniformity fixes, the residual small-$\ell$ gap being the
$\mathcal O((M\ell)^{-1})$ tail. Panel~(b) sweeps the anisotropy at fixed $\ell M=8$, versus $\beta^2$
(top axis $\Delta_z$): the lines carry the two-kink$+B_1$ coefficient $\mathfrak c(\beta^2)$ of
Eq.~\eqref{eq.final}, the replica branches offset rigidly by the universal constant, while the symbols
are the $\gamma\to0$ iDMRG determinations of Table~\ref{app.tab:grand} and the star marks the
free-fermion point. Together with Fig.~\ref{fig.c}, this maps the intermediate coefficient
$\mathfrak c$ to the physical asymmetry over every coupling and replica index.

\begin{figure}[t]
\includegraphics[width=\columnwidth]{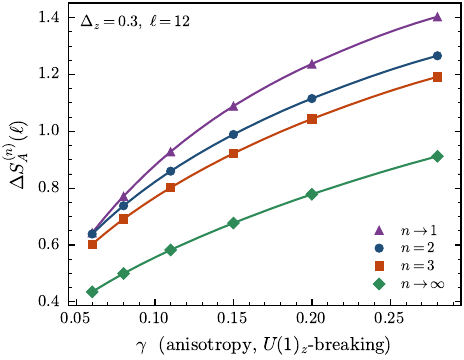}
\caption{$\Delta S^{(n)}_A(\ell)$ for a fixed interval
($\Delta_z=0.3$, $\ell=12$) versus the anisotropy $\gamma$ that controls
the explicit $U(1)_z$ breaking. Colors vary at different R\'enyi indexes $n$.}
\label{fig.gamma}
\end{figure}

Fig.~\ref{fig.gamma} isolates the breaking parameter, tracking $\Delta S^{(n)}_A$ at a fixed
interval as the anisotropy is turned on. As $\gamma\to0$ the chain returns to the symmetric XXZ
point and the asymmetry vanishes; it rises monotonically as $\gamma$ opens the $U(1)_z$ breaking,
ordered by replica index ($n\to1,2,3,\infty$). This is a compact statement that $\Delta S^{(n)}_A$
measures how strongly the ground state breaks the symmetry, and the direct equilibrium counterpart
of the tilted-ferromagnet result of Ref.~\cite{AMC23}, with the anisotropy $\gamma$ in place of the
tilting angle. The values approach the master formula~\eqref{eq.master} once $\ell$ exceeds the
correlation length $1/M$. At fixed $\ell$, the small-$\gamma$ end still carries the
$\mathcal O((M\ell)^{-1})$ crossover, the lattice echo of the non-commuting $\ell\to\infty$ and
$\gamma\to0$ limits.

\section{Discussion}
\label{sec:discussion}
For the gapped $U(1)$-breaking XYZ chain we have obtained a universal susceptibility formula for the
entanglement asymmetry, a two-kink$+B_1$ lower bound for its scaling amplitude
$\mathfrak c(\beta^2)$, two moment sum rules that confine the multi-particle remainder to a two-sided
corridor on the attractive side, the tangent $\mathfrak c'(1)=\pi/2-1$, and the leading
near-free-fermion (four-kink) correction, which we bound from below and estimate perturbatively. To our
knowledge this is the first such analytic characterization for an interacting model with an explicitly
broken $U(1)$ at equilibrium. The amplitude is thereby pinned across the phase, from both sides for
$\xi_{\rm sG}<1$, and completed by the four-kink term that accounts for the small excess seen in iDMRG.

The two halves of the result differ in reach. The master formula needs only extensive, Gaussian
charge fluctuations and exponential clustering, and the sine-Gordon input enters it through $\chi_{zz}$
alone; we expect it to hold, with the corresponding susceptibility, in other gapped chains with an
explicitly broken $U(1)$. The lower bound, the corridor and the tangent law rely on the sine-Gordon form
factors and on the bosonized form of the charge, and are specific to the scaling limit of the XYZ chain.

The breaking vertex saturates a Seiberg-type normalizability bound~\cite{Seiberg90,LZ97} at the
free-fermion point: the LZ vacuum expectation value $\mathcal G_\beta$ diverges for all
$\xi_{\rm sG}\ge1$, and the equation of motion identity used above supplies its continuation, so the
marginality at $\beta^2 = 1/2$ reappears here in a new form, and a third time as the logarithmic
divergence of the $f$-sum rule at $\beta^2 = 1/2$ [Appendix~\ref{app.sec:fourkink}]. The same marginality
controls the tail: at the free-fermion point, the source $\sin\beta\phi$ collapses to the free Dirac mass
$\bar\psi i\gamma^5\psi$, so every $\ge3$-particle channel switches off and the excess is forced to
vanish as $(\xi_{\rm sG}-1)^2$.

Open directions include the finite-$\xi_{\rm sG}$ off-shell-Bethe resummation of the four-kink tail
beyond its computed leading term, a convergent repulsive-side counterpart of the $f$-sum rule that would
close the corridor for $\xi_{\rm sG}>1$, the N\'eel-driven regime $\Delta_z>1$ (where the breaking is
staggered and the operator sits at the same Seiberg bound), finite fields, and post-quench dynamics,
for which Eq.~\eqref{eq.final} sets the equilibrium baseline for Mpemba-type behavior~\cite{RVC24}.

\medskip
\noindent\textit{Data availability.} Codes and data used in this work are provided at \href{https://github.com/ftahas/Entanglement_Asymmetry_XYZ}{https://github.com/ftahas/Entanglement\_Asymmetry\_XYZ}.

\appendix

\section{Master formula}
\subsection{Charged moments}\label{app.sec.charged_moments}
Considering $Q_A$ integer-spaced on an even interval, $\rho_{A,Q}=\sum_q\Pi_q\rho_A\Pi_q$ gives~\cite{AMC23,MAKC23}
\begin{equation}
    \Tr\rho_{A,Q}^2=\frac1{2\pi}\int_{-\pi}^{\pi}d\alpha\,\Tr[\rho_A e^{i\alpha Q_A}\rho_A e^{-i\alpha Q_A}].
\end{equation}
In the eigenbasis $\rho_A|n\rangle=p_n|n\rangle$, we expand it to $\mathcal O(\alpha^2)$, yielding
\[
\begin{split}
\Tr[\rho e^{i\alpha Q}\rho e^{-i\alpha Q}]&=\Tr\rho^2\\
&-\alpha^2\Big[\sum_n p_n^2\left(\langle n|Q^2|n\rangle-Q_{nn}^2\right)\\
&\qquad-\sum_{n\neq m}p_np_m|Q_{nm}|^2\Big]+\mathcal O(\alpha^4).
\end{split}
\]
Using $\langle n|Q^2|n\rangle-Q_{nn}^2=\sum_{m\neq n}|Q_{nm}|^2$ and symmetrizing $n\leftrightarrow m$, we have
\begin{equation}
\begin{aligned}
Z_2(\alpha)&\equiv\frac{\Tr[\rho e^{i\alpha Q}\rho e^{-i\alpha Q}]}{\Tr\rho^2}\\
&=1-V\alpha^2+\mathcal O(\alpha^4),\\
V&=\frac{1}{2\Tr\rho^2}\sum_{n\neq m}|Q_{nm}|^2(p_n-p_m)^2.
\end{aligned}
\label{app.eq.V}
\end{equation}
Observe that it vanishes if and only if $[Q,\rho]=0$, implying that
$\Delta S^{(2)}=0$ for symmetric states. When $V=v\ell$ is extensive and $Z_2$ Gaussian-dominated, we get 
\begin{equation}
    \Delta S^{(2)}=\frac12\log(4\pi V)
     .
\end{equation}

For a fermionic Gaussian $\rho$ diagonal in modes $d_k$
(occupations $f_k$) and a one-body $Q=\sum A_{kl}d^\dagger_kd_l+\frac12(B_{kl}d^\dagger_kd^\dagger_l
+{\rm h.c.})$, Wick evaluation of Eq.~\eqref{app.eq.V} [using $\rho^{-1}d^\dagger_k\rho=e^{\epsilon_k}
d^\dagger_k$ and that $\rho^2/\Tr\rho^2$ is Gaussian with $\tilde f_k=f_k^2/D_k$,
$D_k=f_k^2+(1-f_k)^2$] gives the mode sums
\begin{equation}
\begin{aligned}
V=&\frac12\sum_{kl}|A_{kl}|^2\frac{(f_k-f_l)^2}{D_kD_l}\\
&+\sum_{k<l}|B_{kl}|^2\frac{\left[f_kf_l-(1-f_k)(1-f_l)\right]^2}{D_kD_l},\\
\mathrm{Var}(Q)=&\frac12\sum_{kl}|A_{kl}|^2\left[f_k(1-f_l)+f_l(1-f_k)\right]\\
&+\sum_{k<l}|B_{kl}|^2\left[f_kf_l+(1-f_k)(1-f_l)\right].
\end{aligned}
\end{equation}

The weights coincide whenever both occupations are sharp, $f\in\{0,1\}$. 
For the reduced density matrix of a gapped chain the
single-particle entanglement spectrum has $\mathcal O(1)$ level spacing [as follows, e.g., from the
corner-transfer-matrix structure of the gapped chain~\cite{Baxter,EEFR12,EEFR11,EER10}], so the number of non-sharp modes
is $\mathcal O(1)$ uniformly in $\ell$; since $(A^2)_{kk}\le\|A\|^2\le1$ bounds each mode's contribution,
$V-\mathrm{Var}(Q_A)=\mathcal O(1)$ against the extensive common part:
$V=\mathrm{Var}(Q_A)[1+\mathcal O(1/\ell)]$ for Gaussian states.

Let $|\psi\rangle$ be an injective infinite matrix-product state (MPS) with a $u$-site unit cell, right-canonical tensors $B$,
and a transfer matrix with nondegenerate leading eigenvalue (the variational class of all our
simulations). The statements below are theorems for injective MPS of
finite bond dimension. Their application to the XYZ ground state rests on the fact that a gapped,
exponentially clustering one-dimensional ground state is approximated by this class with an error that
decreases rapidly in the bond dimension.
Define the phase-dressed doubled transfer operator per cell,
\begin{equation}
\begin{aligned}
\mathcal T_\alpha&=\sum_{\vec s\in\rm cell} e^{-i\alpha m_{\vec s}}
B^{\vec s}\otimes\bar B^{\vec s},\\
\lambda(\alpha)&=\text{leading eigenvalue},\quad \lambda(0)=1,
\end{aligned}
\end{equation}
with $m_{\vec s}$ the $Q$-charge of the cell configuration. For a bulk interval of $\ell=uL$ sites:
(i) the characteristic function is a boundary contraction of $\mathcal T_{-\alpha}^{L}$, so
$\log \langle e^{i\alpha Q_A}\rangle=L\log \lambda(-\alpha)+\log b(\alpha)+\mathcal O(e^{-L/\xi_T})$ with $b$ a smooth
boundary overlap; since the layer swap $S(X\otimes Y)S=Y\otimes X$ gives
$S\mathcal T_{-\alpha}S=\bar{\mathcal T}_\alpha$, the spectra of $\mathcal T_{-\alpha}$ and
$\mathcal T_\alpha$ are complex conjugate, and the second cumulant per cell is
$\kappa_2/L=-\partial_\alpha^2\log |\lambda(\alpha)|\big|_0$.
(ii) The charged second moment is
$Z_2(\alpha)=\sum_{ac}\lambda_a^2\lambda_c^2\sum_{bd}|(\mathcal T_\alpha^{L})_{(ac),(bd)}|^2$, so
$\log [Z_2(\alpha)/Z_2(0)]=2L\log |\lambda(\alpha)|+\log [C(\alpha)/C(0)]+\mathcal O(e^{-L/\xi_T})$ with $C$ a smooth
boundary factor. Hence
\begin{equation}
\begin{aligned}
V&=-\frac12\partial_\alpha^2\log Z_2\big|_0\\
&=-L\partial_\alpha^2\log |\lambda(\alpha)|\big|_0+\mathcal O(1)
=\mathrm{Var}(Q_A)+\mathcal O(1)\\
&\Longrightarrow\;\;V=\mathrm{Var}(Q_A)[1+\mathcal O(1/\ell)].
\end{aligned}
\end{equation}

\subsection{General R\'enyi index}\label{app.sec.general_renyi}
Inserting $\Pi_q=\int\frac{d\alpha}{2\pi}e^{i\alpha(Q-q)}$ for each projector,
$\Tr\rho_{A,Q}^n=\int\prod_j\frac{d\alpha_j}{2\pi}
\Tr\left[\rho e^{iu_1Q}\rho e^{iu_2Q}\cdots\rho e^{iu_nQ}\right]$, $u_j=\alpha_j-\alpha_{j+1}$.
Expanding to $\mathcal O(u^2)$ and symmetrizing as in Eq.~\eqref{app.eq.V} yields the replica-resolved widths
\begin{equation}
\begin{aligned}
V_d&=\frac{1}{2\Tr\rho^n}\sum_{n\neq m}|Q_{nm}|^2(p_n^d-p_m^d)(p_n^{n-d}-p_m^{n-d}),\\
&\qquad\qquad\qquad\qquad\qquad d=1,\dots,n-1,
\end{aligned}
\label{app.eq.Vd}
\end{equation}
with $V_d=V_{n-d}$; at $n=2$ this is Eq.~\eqref{app.eq.V}. If $V_d=V$ for all $d$ (the reflection
symmetry makes this automatic for $n\le3$; for general $n$ it is proven below),
then $\log Z_n=-\frac V2\sum_j(\alpha_j-\alpha_{j+1})^2$ and the toroidal Gaussian integral over
the cycle-graph Laplacian [eigenvalues $4\sin^2(\pi k/n)$, $\prod_{k=1}^{n-1}2\sin(\pi k/n)=n$,
zero-mode length $2\pi\sqrt n$] yields
\begin{equation}
\begin{aligned}
\frac{\Tr\rho_{A,Q}^n}{\Tr\rho_A^n}&=\frac{1}{\sqrt n}(2\pi V)^{-(n-1)/2}\\
\Longrightarrow\quad\Delta S^{(n)}_A&=\frac12\log (2\pi V)+\frac{\log n}{2(n-1)}\\
&=\frac12\log (\pi\ell\chi_{zz})+\frac{\log n}{2(n-1)}.
\end{aligned}
\end{equation}
The von Neumann limit $n\to1$, which can be performed through analytic continuation, results in $\frac12\log (\pi e\ell\chi_{zz})$.

Let us write the $n$-sheet charged partition function
\begin{equation}
\begin{aligned}
Z_n(\vec u)&\equiv\frac{\Tr[\rho_Ae^{iu_1Q}\rho_Ae^{iu_2Q}\cdots\rho_Ae^{iu_nQ}]}{\Tr\rho_A^n},\\
&\qquad\qquad\qquad\qquad \sum_{r=1}^{n}u_r=0,
\end{aligned}
\label{app.eq.Znu}
\end{equation}
the zero-sum constraint following from $u_r=\alpha_r-\alpha_{r+1}$ on the replica ring. In the eigenbasis
$\rho_A|a\rangle=p_a|a\rangle$, its second-order expansion is
\begin{equation}
\log Z_n=-\frac12\sum_{r\neq r'}u_ru_{r'}\Phi_{|r-r'|}-\frac12\left(\sum_r u_r^2\right)\Phi_0 ,    
\end{equation}
with
\begin{equation}
\begin{aligned}
\Phi_d&=\frac{1}{\Tr\rho^n}\sum_{a\neq b}|Q_{ab}|^2p_a^{d}p_b^{n-d},\\
V_d&=E-\Phi_d,\quad E\equiv\frac{1}{\Tr\rho^n}\sum_{a\neq b}|Q_{ab}|^2 p_a^{n},
\end{aligned}
\label{app.eq.VdPhi}
\end{equation}
such that $\Phi_d=-\langle Q_rQ_{r+d}\rangle_c^{\rm off}$ is (minus) the off-diagonal connected charge--charge 
correlator at replica separation $d$, and the $d$-independence of $E$ reduces the claim $V_d=V$ to the
$d$-independence of the extensive part of $\Phi_d$.

We compute $Z_n(\vec u)$ with the phase-dressed doubled transfer operator, taken below in the conjugate phase convention $\mathcal T_u$,
Eq.~\eqref{app.eq.Tu}. The two spectra are complex conjugate (Sec.~\ref{app.sec.charged_moments}), so the
second-order expansion is unaffected by the choice. For a bulk interval of $\ell=uL$ cells, the phase $e^{iu_rQ}$ dresses the $r$-th ket
layer. Gluing the $n$ layers cyclically, as the trace in Eq.~\eqref{app.eq.Znu} demands, we factorize the
per-cell operator into a tensor product of $\mathcal T_{u_r}$, whence
\begin{equation}
Z_n(\vec u)\Tr\rho_A^n
=\left\langle\mathcal B_L\left|\textstyle\bigotimes_{r=1}^{n}\mathcal T_{u_r}^{L}\right|\mathcal B_R\right\rangle,
\end{equation}
with $\ell$-independent boundary tensors $\mathcal B_{L,R}$ built from the Schmidt weights.

The factorization deserves explicit indices, as it is the central step. We now index the ket virtual bond of
replica $r$ by $a_r$ and the bra virtual bond by $b_r$. Writing each $\rho_A$ in right-canonical form
and inserting $e^{iu_rQ}=\prod_c e^{iu_r m_{s_c}}$ on the $r$-th ket layer, the bulk contraction is, per
cell, the tensor product
\begin{equation}
\bigotimes_{r=1}^{n}\left(\mathcal T_{u_r}\right)_{(a_rb_r),(a_r'b_r')},\qquad
\mathcal T_{u}=\sum_{s}e^{ium_{s}}B^{s}\otimes\bar B^{s},
\label{app.eq.Tu}
\end{equation}
with each factor acting on its own doubled bond space $(a_rb_r)$. The cyclic trace of
Eq.~\eqref{app.eq.Znu} glues the bra of sheet $r$ to the ket of sheet $r{+}1$. In the doubled variables, it corresponds to the cyclic permutation $\pi:b_r\mapsto a_{r+1}$ of the bra bonds relative to the ket bonds. Since
the bulk chain is uniform, $\pi$ commutes with $\bigotimes_r\mathcal T_{u_r}^{L}$ except at the two
interval ends, where it is absorbed into $\mathcal B_{L,R}$. The entire replica coupling therefore
resides in the boundary tensors---hence in the $\mathcal O(1)$ boundary Hessian below---while the
extensive part is the replica-diagonal $\bigotimes_r\mathcal T_{u_r}^{L}$.  Injectivity gives each $\mathcal T_u$ a nondegenerate leading eigenvalue $\lambda(u)$ ($\lambda(0)=1$, analytic near
$u=0$) separated by a gap, so $\mathcal T_u^{L}=\lambda(u)^L|R(u)\rangle\langle L(u)|+\mathcal O(e^{-L/\xi_T})$ and
\begin{equation}
\log Z_n(\vec u)=L\sum_{r=1}^{n}\log \lambda(u_r)+\log \mathcal B(\vec u)+\mathcal O(e^{-L/\xi_T}),
\label{app.eq.lnZn}
\end{equation}
with $\mathcal B(\vec u)$ a smooth boundary overlap.
Two features of Eq.~\eqref{app.eq.lnZn} settle the proof. Firstly, the extensive term is
separable---a sum of single-variable functions of the individual $u_r$---so its mixed second
derivatives $\partial_{u_r}\partial_{u_{r'}}$ with $r\neq r'$ vanish. Then, the linear
term $iL\,\mathrm{Im}\lambda'(0)\sum_r u_r$ vanishes by $\sum_r u_r=0$. Writing
$-\partial_u^2\log \lambda|_0=\mathrm{Var}(Q_A)/L$ for the second-cumulant density,
\begin{equation}
\begin{aligned}
\log Z_n(\vec u)=&-\frac{\mathrm{Var}(Q_A)}{2}\sum_{r=1}^{n}u_r^2\\
&+\frac12\vec u^{\mathrm T}\left[\partial^2\log \mathcal B\right]_0\vec u+\mathcal O(u^3),
\end{aligned}
\end{equation}
and $\sum_r u_r^2=\sum_r(\alpha_r-\alpha_{r+1})^2=\vec\alpha^{\mathrm T}\mathbb L_n\vec\alpha$ is the
cycle-graph Laplacian form. The extensive part therefore carries the single coefficient
$\mathrm{Var}(Q_A)$ and no longer-range couplings; the entire $d$-dependence of the quadratic form
resides in the $\mathcal O(1)$ boundary Hessian $\partial^2\log \mathcal B$. By Eq.~\eqref{app.eq.VdPhi} this means
$\Phi_d^{\rm ext}$, hence $V_d^{\rm ext}$, is independent of $d$. Consequently, 
\begin{equation}
\begin{aligned}
V_d&=\mathrm{Var}(Q_A)[1+\mathcal O(1/\ell)]\\
&=\frac{\ell\chi_{zz}}{2}[1+\mathcal O(1/\ell)],\qquad d=1,\dots,n-1,
\end{aligned}
\end{equation}
for every $n$, the $d$-dependent remainder corresponding to the boundary curvature to which $V_d-V$
saturates.  

We confirm the mechanism at the free-fermion point
($\Delta_z=0$, the XY line), where the interval reduced density matrix is Gaussian and the widths~\eqref{app.eq.Vd} are obtained. 
All $V_d^{(n)}$ grow with the common slope $\chi_{zz}/2$ while the differences
$V_d-V_1$ saturate to $\ell$-independent constants with reflection symmetry $V_d=V_{n-d}$. 
Such saturation constants correspond to $\mathcal O(1)$
boundary curvatures $\partial^2\log \mathcal B$ of the proof.

\section{Sum rule and form factors}
\subsection{Sum rule}\label{app.sec:sumrule}
On the lattice, where $Q_A=\frac12\sum_{j\in A}\sigma^z_j$, the static
susceptibility is the connected structure factor per site, 
$\chi_{zz}=\frac12\sum_r\langle\sigma^z_0\sigma^z_r\rangle_c=2\lim_{\ell\to\infty}\mathrm{Var}(Q_A)/\ell$.
In the sine-Gordon scaling limit, the uniform part of $\sigma^+\sigma^+$ is the vertex $e^{i\beta\phi}$ of
$U(1)_z$ charge $q=2$. Matching the lattice generator $Q=\frac12\sum_j\sigma^z_j$ to the continuum gives
$Q=(q/\beta)\int dx\,\Pi$ with $\Pi=\partial_t\phi/4\pi$ and $q=2$. It means that the normalization of 
the $U(1)_z$ current is fixed at the XXZ point, where $Q$ generates the
symmetry. The $\gamma$-perturbation renormalizes only the breaking vertex and leaves the conserved-current normalization, hence $q=2$, untouched. We use LZ conventions~\cite{LZ97},
$\mathcal A=\int d^2x\,[\frac1{16\pi}(\partial\phi)^2-2\mu\cos\beta\phi]$, with rapidity states
normalized as $\langle\theta|\theta'\rangle=4\pi\delta(\theta-\theta')$.

The non-conservation of $Q$, Eq.~\eqref{eq.Qdot}, gives the spectral sum rule~\eqref{app.eq.spectral}
of Sec.~\ref{sec:spectral}, which we evaluate channel by channel.

A two-kink state $|\theta_1\theta_2\rangle$ has $E=M(\cosh\theta_1+\cosh\theta_2)$ and
$\langle\theta_1\theta_2|\sin\beta\phi|0\rangle\propto F^{\sin}(\theta_{12})\delta_{\rm c.m.}$. The
center-of-mass momentum delta cancels one power of $L$ against the $1/L$ in Eq.~\eqref{app.eq.spectral} and
sets $\theta_1=-\theta_2$, resulting in 
\begin{equation}
\chi_{2\rm k}=\frac{q^2 M}{16\pi\beta^4}\int_{-\infty}^{\infty}d\theta\,
\frac{|\mathcal F(2\theta)|^2}{\cosh\theta},
\label{app.eq.chi2k}
\end{equation}
where $F^{\sin}(\theta_{12})=\frac{M^2}{2\mu\beta^2}\cosh\frac{\theta_{12}}2\mathcal F(\theta_{12})$. 
The $B_1$ term is the isolated one-particle pole
$\chi_{B_1}=2(2q\mu)^2|F^{\sin}_{B_1}|^2/(2m_1^3)$, with mass $m_1=2M\sin(\pi\xi_{\rm sG}/2)$. 
The single constant prefactor in Eq.~\eqref{app.eq.chi2k} and in $\chi_{B_1}$ is pinned by the free-fermion
point [check~(a) below].

Lukyanov's two-soliton form factors of $e^{ia\beta\phi}$~\cite{Luk97} combine in the
$C$-odd channel to $F^{\sin}\propto \mathcal G_\beta\cot(\pi\xi/2)[G(\theta)/G(-i\pi)]\cosh(\theta/2)/
\cosh\frac{\theta+i\pi}{2\xi}$. The equation of motion,
$\sin\beta\phi=-\Box\phi/(16\pi\mu\beta)$, applied to his field form factor fixes the constant through
\begin{equation}
\mathcal G_\beta\cot\frac{\pi\xi}{2}=\frac{M^2\xi}{8\mu\beta^2}.
\label{app.eq.EOM}
\end{equation}
This last identity comes from the Hellmann--Feynman theorem
applied to the action, $\mathcal G_\beta=\langle\cos\beta\phi\rangle=-f/[2(1-\beta^2)\mu]$ with $f$ the
ground-state energy density. Using $M\propto\mu^{1/(2-2\beta^2)}$~\cite{Zam95}
and $\xi(1-\beta^2)/\beta^2=1$, Eq.~\eqref{app.eq.EOM} is equivalent to the known sine-Gordon bulk energy
\begin{equation}
f=-\frac{M^2}{4}\tan\frac{\pi\xi}{2}
\end{equation}
\cite{Zam95,DdV95}. We additionally verified
Eq.~\eqref{app.eq.EOM} against the LZ vacuum-expectation-value (VEV) integral~\cite{LZ97} on the attractive
side---the VEV integral converges only for $\xi<1$: at $a=\beta$ the operator saturates a Seiberg-type bound~\cite{Seiberg90} at the free-fermion point, and Eq.~\eqref{app.eq.EOM} provides the repulsive-side continuation. The result is
\begin{equation}
\begin{aligned}
\mathcal F(\theta)&=\frac{\sinh(\theta/2)e^{\mathcal J(\theta)}}{\cosh\frac{\theta+i\pi}{2\xi}},\\
\mathcal J&=-\int_0^\infty\frac{dt}{t}\frac{k_0(t;\xi)}{\sinh t}
\sin^2\frac{t(i\pi+\theta)}{2\pi},\\
k_0&=\frac{\sinh\frac{(\xi-1)t}2}{\sinh\frac{\xi t}2\cosh\frac t2}.
\end{aligned}
\label{app.eq.Ftrue}
\end{equation}
The following checks hold. (a) Free-fermion point: $\mathcal F(\theta;1)\equiv1$ analytically, and Eq.~\eqref{app.eq.chi2k} integrates to
$\chi=M$ ($\mathfrak c(\frac12)=1$), matching the XY value $\chi=\gamma/(1+\gamma)$,
$M=\mathrm{atanh}\gamma$, in the scaling limit; (b) Watson's equation with the $C$-odd eigenvalue
$(c-b)$; (c) large-$\theta$ growth $e^{(\xi-1)|\theta|/4\xi}$
(so $|F^{\sin}|\sim e^{(3\xi-1)|\theta|/4\xi}$, below the ultraviolet (UV) bound $e^{\beta^2|\theta|}$ with margin
$(\xi-1)^2/4\xi(\xi+1)$, saturating only at $\xi=1$); (d) the representation \eqref{app.eq.Ftrue} coincides
numerically, at all tested $\theta$ and $\xi$, with an independently constructed solution built on the
Karowski--Weisz kernel~\cite{KW78} ($\Delta h=2-2\sinh t/\sinh\xi t-k_0$, argument
$i\pi-\theta$, normalized at $i\pi$)---a nontrivial identity between two integral representations.
Check (c) is also a minimality statement: multiplying $\mathcal F$ by a nontrivial Castillejo--Dalitz--Dyson (CDD) factor
either violates the UV bound---whose margin closes only at the free-fermion point---or introduces
poles absent from the two-kink kinematics; check (d) confirms nonperturbatively that the two
integral representations define the same function.

Lukyanov's one-breather matrix element~\cite{Luk97} yields
\begin{equation}
\begin{aligned}
F^{\sin}_{B_1}&=\sqrt2\lambda(\xi)\mathcal G_\beta , \\
\lambda(\xi)&=2\cos\frac{\pi\xi}2\sqrt{2\sin\frac{\pi\xi}2}\;
e^{-\frac1{2\pi}\int_0^{\pi\xi}t\,dt/\sin t}.
\end{aligned}
\end{equation}
Eliminating $\mathcal G_\beta$ via \eqref{app.eq.EOM} and using the breather mass $m_1$ above, we have
\begin{equation}
\begin{aligned}
\mathfrak c_{B_1}(\xi)&=\frac{\chi_{B_1}}{M}
=\frac{\xi^2}{2\beta^4}\exp\left(-\frac1\pi\int_0^{\pi\xi}\frac{t\,dt}{\sin t}\right)\\
&=\frac{(1+\xi)^{2}}{2}\left(\tan\frac{\pi\xi}{2}\right)^{-\xi}\\
&\quad\times\exp\left[\frac{\mathrm{Cl}_2(2\pi\xi)}{2\pi}-\frac{2\mathrm{Cl}_2(\pi\xi)}{\pi}\right],
\end{aligned}
\label{app.eq.cB1closed}
\end{equation}
where $\mathrm{Cl}_2(\pi\xi)$ is the Clausen function $\mathrm{Cl}_2(x)=\sum_{k\ge1}\frac{\sin kx}{k^2}$, useful here as the solution 
of $\int_0^{x}\frac{t\,dt}{\sin t}=x\log\tan\frac{x}{2}+2\mathrm{Cl}_2(x)-\frac12\mathrm{Cl}_2(2x)$.
Some convenient consequences are in place. Firstly,  $\mathfrak c_{B_1}(\xi\to0)=\frac12$ (the
$\xi\to0$ amplitude is carried entirely by $B_1$),
$\mathfrak c_{B_1}(\frac12) \propto e^{-2G/\pi}$ with $G$ Catalan's constant, and
\begin{equation}
\mathfrak c_{B_1}(\xi\to1^-)=\pi(1-\xi)+O\left((1-\xi)^2\right),
\label{app.eq.B1linear}
\end{equation}
the linear vanishing that enters the tangent law at the free-fermion point. The closed form
agrees with the fusion-residue route (residue of \eqref{app.eq.Ftrue} at $\theta=-i\pi(1-\xi)$ with
the coupling $g^2=2\xi\cos\frac{\pi(2-\xi)}{2\xi}S_0(i\pi(1-\xi))$). Since the sum rule is a sum of squares and the form factors are the minimal
solutions, $\mathfrak c_{2\rm k}+\mathfrak c_{B_1}$
composes a lower bound on $\mathfrak c$. Its leading correction---the four-kink channel---is estimated in Sec.~\ref{app.sec:fourkink} to leading order near the free-fermion point: it vanishes at the free-fermion
point and enters as $\mathfrak c_{4\rm k}=A(\xi-1)^2+\dots$ with $A\simeq0.08$ from a Monte-Carlo evaluation, an estimate of the multi-particle tail near the free-fermion point. 
The bound values in Table~\ref{app.tab:grand} were evaluated with two independent quadratures (fixed Gauss--Legendre panels
resolving the oscillatory $\mathcal J$ integral, and adaptive quadrature with an oscillatory-weight rule), which agree to high precision; for $\xi>1$ the integrand of Eq.~\eqref{app.eq.chi2k} decays only as
$e^{-\theta/\xi}$, so large rapidity cutoffs are needed to reach full accuracy.

\subsection{The four-kink tail}\label{app.sec:fourkink}
The excess of $\mathfrak c$ over $\mathfrak c_{2\rm k}+\mathfrak c_{B_1}$ is the weight of the
$\ge3$-particle intermediate states in the sum rule. On the repulsive side ($\xi>1$) there are no
breathers, so the leading contribution is the four-kink (two-soliton--two-antisoliton) state. We establish that it switches on quadratically at the free-fermion point, and give a leading-order estimate of its coefficient.

At $\xi=1$ ($\beta^2=\frac12$) the sine-Gordon model is a free massive Dirac fermion, and the
source operator is the fermion mass bilinear,
\begin{equation}
\sin\beta\phi\;\xrightarrow{\ \xi\to1\ }\;\frac{M}{4\mu\beta^2}\bar\psi i\gamma^5\psi,
\label{app.eq.coleman}
\end{equation}
the normalization fixed by matching the two-particle form factor $F^{\sin}$ 
(which reproduces $\mathfrak c_{2\rm k}(1)=1$). A free-field bilinear has form factors only in the
one-particle--one-antiparticle sector; its connected four-particle form factor vanishes. Hence $\mathfrak c_{4\rm k}(\xi{=}1)=0$, and since the leading connected four-particle form
factor is generated at first order in the Thirring coupling $g$, with $g=-\frac{\pi}{2}(\xi-1)+\mathcal O((\xi-1)^2)$, one has $|F^{\sin}_4|^2=\mathcal O(g^2)$ and
\begin{equation}
\mathfrak c_{4\rm k}(\xi)=A(\xi-1)^2+O\left((\xi-1)^3\right),
\label{app.eq.c4k}
\end{equation}
corroborating the $\mathcal O((\xi-1)^2)$ law asserted above, and identifying its microscopic origin: the
breaking source degenerates to a free mass term at the free-fermion point.

The order-$g$ connected four-particle form factor of $\bar\psi i\gamma^5\psi$
is the single-vertex Thirring amplitude ($-\frac{g}{2}(\bar\psi\gamma^\mu\psi)^2$), obtained by Wick
contraction with one internal fermion propagator carrying the three-particle invariant momentum
($k^2\ge9M^2$); it is purely imaginary (the operator is $C$-odd), boost invariant
(a Lorentz pseudoscalar), and Fermi-antisymmetric under exchange of like charges. Feeding it into
Eq.~\eqref{app.eq.spectral} with the measure of Sec.~\ref{app.sec:sumrule}, at zero total momentum
$\sum_i\sinh\theta_i=0$ and $E=M\sum_i\cosh\theta_i$,
\begin{equation}
\begin{aligned}
A&=\frac{\pi^2}{12}J,\\
J&=\frac{6}{128\pi^3}\int d\theta_1 d\theta_2 d\theta_3\,
\frac{\big|f_{++--}(\vec\theta)\big|^2}{(\textstyle\sum_i\cosh\theta_i)^2\cosh\theta_4},
\end{aligned}
\label{app.eq.Aint}
\end{equation}
$f_{++--}$ the amplitude~\eqref{app.eq.coleman} stripped of $g$, and $\theta_4$ fixed by momentum
conservation. The integrand is ultraviolet convergent (a hard rapidity cutoff saturates once the span is
a few units wide), and Monte-Carlo quadrature gives $A\simeq0.08$, the four like-charge configurations
contributing equally. This is a leading-order coefficient (first order in $\xi_{\rm sG}-1$) obtained by
numerical integration; independently of any Monte-Carlo input, the moment sum rules bound it from below.
The explicit closed form of $f_{++--}$ with its Lehmann--Symanzik--Zimmermann (LSZ) and external-state normalizations, the enumeration of
contractions and crossing assignments, and the disconnected subtractions are provided with the code in the repository; $A$ is to be read as a perturbative estimate near the free-fermion point.

Close to the free-fermion point, such a law accounts for the small excess of the measured coefficient over the
two-kink$+B_1$ value: the $(\xi_{\rm sG}-1)^2$ onset is borne out, and the four-kink line matches the
largest repulsive coupling $\xi_{\rm sG}=2$. Away from the free-fermion point the leading-order line overshoots---at
the intermediate repulsive couplings the measured excess is consistent with the bound but sits below the
positive $A(\xi_{\rm sG}-1)^2$ prediction---as expected for a leading term in $\xi_{\rm sG}-1$, since the
omitted finite-$\xi_{\rm sG}$ off-shell dressing of the four-kink form factor is not negligible there. On
the attractive side ($\xi_{\rm sG}<1$) the leading $\ge3$-particle channels involve breathers
($s\bar sB_1,\dots$). The multi-particle tail is thus non-negative, bounded below by the
two-kink$+B_1$ value, resolved in the data near $\xi_{\rm sG}=2$, captured near the free-fermion point by the
four-kink law, and confined by two moment sum rules to a two-sided
corridor for $\xi_{\rm sG}<1$ and bounded from below for $\xi_{\rm sG}>1$, such that
\begin{equation}
\mathfrak c(\beta^2)=\mathfrak c_{2\rm k}(\xi_{\rm sG})+\mathfrak c_{B_1}(\xi_{\rm sG})\Theta(1-\xi_{\rm sG})+A(\xi_{\rm sG}-1)^2+\dots
\end{equation}
is the lower bound completed, near the free-fermion point, by the leading tail: the first two terms in
closed form [Eq.~\eqref{app.eq.cB1closed}], the third a perturbative estimate near the free-fermion point.

We now consider the sum rule of Sec.~\ref{app.sec:sumrule} as a family of spectral moments,
\begin{equation}
T_j\equiv\frac{2}{L}\sum_{n\neq0}E_n^{j}\big|\langle n|Q|0\rangle\big|^2,
\end{equation}
so that $T_0=\chi_{zz}=M\mathfrak c$ is Eq.~\eqref{app.eq.spectral}, and the two-kink and $B_1$
channels contribute, with the measure of Eq.~\eqref{app.eq.chi2k},
\begin{equation}
\begin{aligned}
T_j^{2\rm k}&=\frac{q^2M^{1+j}}{16\pi\beta^4}\int_{-\infty}^{\infty}d\theta\,
\frac{|\mathcal F(2\theta)|^2}{\cosh\theta}(2\cosh\theta)^{j},\\
T_j^{B_1}&=M\mathfrak c_{B_1}m_1^{j}.
\end{aligned}
\label{app.eq.Tj2k}
\end{equation}

Couple a uniform field to the broken
charge, $H(h)=H-hQ$, $Q=(q/\beta)\int dx\,\Pi$. Because $Q$ is built from the field momentum
alone, completing the square and conjugating by $e^{i\lambda\int\phi}$ (which shifts
$\Pi\to\Pi+\lambda$ and commutes with every $\phi$-dependent term of $H$) removes the perturbation,
\begin{equation}
E_0(h)=E_0(0)-\frac{h^2q^2}{16\pi\beta^2}L\qquad\text{to all orders in }h,
\end{equation}
so second-order perturbation theory evaluates the static Kubo susceptibility:
\begin{equation}
T_{-1}=\frac{2}{L}\sum_{n\neq0}\frac{|\langle n|Q|0\rangle|^2}{E_n}
=\frac{q^2}{8\pi\beta^2}=\frac{1+\xi}{2\pi\xi}.
\label{app.eq.Tm1exact}
\end{equation}
On the lattice, Eq.~\eqref{app.eq.Tm1exact} states that the uniform-field susceptibility of the
gapped chain tends, as $\gamma\to0$, to the XXZ value $K/\pi v$ with $K=1/2\beta^2$: the
spectral gap does not suppress the field response, because the charge is not conserved.

The double commutator
$\frac12\langle[Q,[H,Q]]\rangle=q^2\mu L\mathcal G_\beta$ gives the $f$-sum rule of the
broken charge; eliminating $\mathcal G_\beta$ through Eq.~\eqref{app.eq.EOM},
\begin{equation}
\begin{aligned}
T_{+1}&=\frac{2}{L}\sum_{n\neq0}E_n|\langle n|Q|0\rangle|^2
=2q^2\mu\mathcal G_\beta\\
&=\frac{q^2M^2}{4}(1+\xi)\tan\frac{\pi\xi}{2}\qquad(\xi<1).
\end{aligned}
\label{app.eq.Tp1exact}
\end{equation}
For $\xi\ge1$ the spectral sum is ultraviolet divergent---the $J_0$ integrand grows as
$e^{(\xi-1)|\theta|/\xi}$, cf.\ check (c)---in 
correspondence with the pole of $\tan(\pi\xi/2)$ and with the Seiberg-bound marginality of the
free-fermion point: the growth margin of $|\mathcal F|$ closes where
the $f$-sum rule becomes logarithmically divergent. The finite combination
$r_{+1}\equiv[T_{+1}-T^{2\rm k}_{+1}-T^{B_1}_{+1}]/M^2$ approaches
$r_{+1}(\xi\to1^-)=2/\pi$, a finite limit of a difference of divergent quantities.

The $\ge3$-particle tail defines a positive spectral measure
$d\sigma(E)$ supported on $E\ge E_{\rm th}$, with $E_{\rm th}=4M$ for $\xi>1$ (four-kink) and
$E_{\rm th}=2M+m_1$ for $\frac12<\xi<1$ ($s\bar sB_1$; the $C$-even states $B_1B_1$, $B_2$ do
not couple to $\sin\beta\phi$). Its moments
$r_{-1}=\int d\sigma/E$, $M\mathfrak c_{\rm tail}=\int d\sigma$, $M^2r_{+1}=\int E\,d\sigma$
obey, by Cauchy--Schwarz and by $E\ge E_{\rm th}$ pointwise,
\begin{equation}
\begin{aligned}
E_{\rm th}r_{-1}\;\le\;&\mathfrak c_{\rm tail}\;\le\;\sqrt{r_{+1}r_{-1}}\quad(\xi<1),\\
4r_{-1}\;\le\;&\mathfrak c_{\rm tail}\quad(\xi>1),
\end{aligned}
\label{app.eq.corridor2}
\end{equation}
and both bounds are sharp given the two known moments (the upper bound is attained by a
measure concentrated at $E^*=M\sqrt{r_{+1}/r_{-1}}\ge E_{\rm th}$, the lower by mass at
threshold plus mass escaping to infinity). The moments follow by subtracting the
two-kink$+B_1$ quadratures from the closed forms
\eqref{app.eq.Tm1exact}--\eqref{app.eq.Tp1exact}. The tail moment $r_{-1}$ stays positive at every coupling
computed across the broken phase. Table~\ref{app.tab:grand} lists $r_{-1}$ and the corridor
edges $\mathfrak c_{\rm low/up}$ at the measured couplings. All four attractive determinations fall
inside the corridor, whose width is comparable to the uncertainty of the infinite-system
density-matrix renormalization group (iDMRG) determinations. On the repulsive side the
corridor's lower edge lies above the two central values that sit marginally below the two-kink$+B_1$
bound, which identifies those negative excesses as $\gamma\to0$ extrapolation systematics. At the largest repulsive coupling the bound leaves the tail
consistent with the measured excess and shows that most of the four-kink weight sits at high energy.
Near the free-fermion point the tail moment vanishes quadratically,
$r_{-1}=A_{-1}(\xi-1)^2+O(|\xi-1|^3)$ with $A_{-1}\approx0.0030$ consistently from both sides,
which bounds the four-kink coefficient, i.e. 
\begin{equation}
A\;\ge\;4A_{-1}\;\approx\;0.012.
\label{app.eq.Abound}
\end{equation}

From the representation \eqref{app.eq.Ftrue}, we have that 
\begin{equation}
\partial_\xi\log|\mathcal F(u;\xi)|^2\Big|_{\xi=1}
=1+u\coth\frac u2-\frac u2\tanh\frac u2.
\end{equation}
Integrated against the $J_1$ weight ($u=2\theta$), the elementary integrals
$\int_{\mathbb R}\mathrm{sech}\,\theta\,d\theta=\pi$,
$\int_{\mathbb R}2\theta\coth\theta\mathrm{sech}\,\theta\,d\theta
=2\int_{\mathbb R}\theta\,d\theta/\sinh\theta=\pi^2$ and
$\int_{\mathbb R}\theta\tanh\theta\mathrm{sech}\,\theta\,d\theta=\pi$ give the smooth part
$\partial_\xi J_1|_1=\pi^2$, where
$J_k\equiv\int_{-\infty}^{\infty}d\theta\,|\mathcal F(2\theta)|^2/\cosh^k\theta$
[$\mathfrak c_{2\rm k}=(1+\xi)^2J_1/4\pi\xi^2$]. In addition, the $B_1$ pole of
$1/\cosh\frac{u+i\pi}{2\xi}$ at $u=-i\pi(1-\xi)$ carves a Lorentzian notch of width
$w=\pi|\xi-1|/2$ out of $|\mathcal F(2\theta)|^2$ at threshold
[$|\mathcal F|^2\simeq\theta^2/(\theta^2+w^2)$ locally], removing weight $\pi w=\pi^2|\xi-1|/2$
from $J_1$ on both sides of the free-fermion point. Hence the one-sided derivatives
\begin{equation}
J_1'(1^-)=\frac{3\pi^2}{2},\qquad J_1'(1^+)=\frac{\pi^2}{2},
\end{equation}
and $\mathfrak c_{2\rm k}'(1^+)=\frac\pi2-1$, $\mathfrak c_{2\rm k}'(1^-)=\frac{3\pi}2-1$.
The $-\pi$ slope of $\mathfrak c_{B_1}$, Eq.~\eqref{app.eq.B1linear}, cancels the kink
contribution---the breather weight flows continuously into the two-kink threshold as
$\xi\to1^-$, a further consistency test of Lukyanov's $B_1$ normalization---leaving the full
amplitude differentiable at the free-fermion point with the tangent
\begin{equation}
\mathfrak c(\beta^2)=1+\left(\frac{\pi}{2}-1\right)(\xi-1)+O\left((\xi-1)^2\right).
\label{app.eq.ffpSM}
\end{equation}
An unconstrained fit to the computed $\mathfrak c_{2\rm k}$ on the repulsive side reproduces the 
intercept $1$ and slope $\pi/2-1$. The same computation for $J_2$ gives the smooth part $1+\pi^2/2$ and
one-sided slopes $J_2'(1^+)=1$, $J_2'(1^-)=1+\pi^2$, whence the quadratic onset of $r_{-1}$.

The exponent $2\,\mathrm{Re}\,\mathcal J$ is a
Barnes-$G$-level object (its exponential-series expansion carries coefficients linear in
the index), so $J_1$ is an integral of double-sine-class functions against $\mathrm{sech}$; no member of this family is known to reduce to standard constants. At the reflectionless point
$\xi=\frac12$ [$\beta^2=\frac13$, $\Delta_z=-\frac12$], where $k_0=-1/\cosh(t/2)$ collapses
the exponent to a summable series, $J_0$, $J_1$ and $J_2$ can be evaluated to high
precision. We therefore regard
Eqs.~\eqref{app.eq.cB1closed}, \eqref{app.eq.Tm1exact}, \eqref{app.eq.Tp1exact}, \eqref{app.eq.corridor2},
\eqref{app.eq.Abound} and \eqref{app.eq.ffpSM} as the closed-form content of the sum rule. 

\section{Lattice inputs}\label{app.sec:deriv}
The setup is the following. Mapping $H$ to the rotated principal regime of the eight-vertex model~\cite{Baxter} (via the unitaries that exchange
axes and flip coupling-sign pairs) fixes the eight-vertex couplings, and through them the Boltzmann
anisotropy $\Gamma$ and asymmetry $\Delta$. The standard elliptic parametrization of the principal regime~\cite{EEFR12}
then delivers the Johnson--Krinsky--McCoy (JKM) modulus $k_1'$~\cite{JKM73} with the identities in place
\begin{equation}
\begin{aligned}
J&=1-\gamma,\quad J_y=\frac{1+\gamma}{1-\gamma},\quad J_z=-\frac{\Delta_z}{1-\gamma},\\
\Gamma&=\frac{\Delta_z}{1-\gamma},\quad \Delta=-\frac{1+\gamma}{1-\gamma},
\end{aligned}
\end{equation}
and 
\begin{equation}
\begin{aligned}
l&=\sqrt{\frac{(1-\gamma)^2-\Delta_z^2}{(1+\gamma)^2-\Delta_z^2}},\quad
\mu_{\rm e}=\frac{\pi[K(l')-F(\arcsin\Gamma;l')]}{2K(l')},\\
\epsilon&=\mu_{\rm e}\frac{K(l')}{K(l)},\quad x=e^{-\epsilon},\quad
k_1'=\frac{\theta_4^2(0,x)}{\theta_3^2(0,x)}.
\end{aligned}
\end{equation}
The kink mass is then $M=\mathrm{atanh}\,k_1'$, and it sets the ordered-phase two-particle correlation
length through $\xi_{zz}^{-1}=2M$ for $\Delta_z\ge0$.

At the XY line the Boltzmann anisotropy vanishes, which removes the incomplete elliptic integral and pins
the Poincar\'e parameter to $\mu_{\rm e}=\pi/2$ for every $\gamma$. The nome variable then reduces to the
square root of the nome of $l$, whose modulus is the ascending Landen image of $l$; carrying that Landen
step through and inserting $l=(1-\gamma)/(1+\gamma)$ collapses the JKM modulus to $k_1'=\gamma$, so
that
\begin{equation}
    M=\mathrm{atanh}\,\gamma.
\end{equation}

Regarding the amplitude asymptotic, firstly we fix $\Delta_z$ and let $\gamma\to0^+$, writing $\theta_0=\arccos\Delta_z$. In this limit the complementary
modulus $l'$ vanishes like $\sqrt\gamma$, so its complete integral $K(l')$ tends to $\pi/2$ while $K(l)$
diverges logarithmically. The incomplete integral in the Poincar\'e parameter reduces to its argument,
giving $\mu_{\rm e}\to\theta_0$, and hence a nome exponent $\epsilon\to\pi\theta_0/2L$ controlled by the
single logarithm $L=\frac12\log (1/\gamma)+\log (2\sin\theta_0)$. Passing to the complementary nome by a
modular transformation and expanding $k_1'$ for small nome then produces the essential-singularity form of
the mass,
\begin{equation}
M=\mathrm{atanh}\,k_1'=4e^{-\pi L/\theta_0} 
=4\left(\frac{\sqrt\gamma}{2\sin\theta_0}\right)^{\pi/\theta_0}.
\end{equation}
Raising it to the power $2\theta_0/\pi$ isolates the
amplitude, $M^{2\theta_0/\pi}/\gamma\to 4^{1-2\beta^2}/\sin^2\theta_0$ (verified numerically at
several anisotropies), which with the definition $M=(\gamma\mathcal C/\kappa)^{\pi/2\theta_0}$ fixes
$\mathcal C=\kappa(\beta^2)\,4^{1-2\beta^2}/(1-\Delta_z^2)$.

The last ingredient is the $B_1$ mass and the breather law. 
For $\mu_{\rm e}>\pi/2$, i.e. $\Delta_z<0$, the lattice supports a breather bound state whose mass is
given by the JKM formula~\cite{JKM73}
\[
\begin{aligned}
m_{B_1}&=2\left[\mathrm{atanh}\,k_1'+\log \mathrm{dn}(w;k_2')\right],\\
k_2&=\frac{1-k_1'}{1+k_1'},\qquad
w=\frac{2K(k_2)K(l')(\mu_{\rm e}-\pi/2)}{\pi K(l)},
\end{aligned}
\]
where $\mathrm{dn}(u;k)$ is the Jacobi elliptic delta-amplitude function of modulus $k$. The scaling law follows by taking $\gamma\to0$. In such a limit, the modulus $k_2'$ vanishes, 
so the delta-amplitude logarithm collapses to a small term quadratic in $\sin w$, while the
leading $\mathrm{atanh}$ linearizes, and the mass ratio tends to $\cos2w_\infty$ with $w_\infty$ being the
limiting value of $w$. Evaluating that limit from the small-nome forms of the elliptic integrals gives
\begin{equation}
    w_\infty= \frac{\pi}{4}(1-\xi_{\rm sG}), \qquad \xi_{\rm sG}=\frac{\pi}{\theta_0}-1,
\end{equation}
and hence the breather law 
\begin{equation}
    \frac{m_{B_1}}{2M} \to\sin(\pi\xi_{\rm sG}/2).
\end{equation}

\section{Numerical protocols and data}\label{app.sec:numerics}
\subsection{iDMRG and susceptibility}
We implement TeNPy iDMRG~\cite{tenpy} with the following features: two-site unit cell, N\'eel-$x$ product initial state, no symmetry enforced,
$\chi_{\rm bond}\le160$, truncation $10^{-12}$. The susceptibility
$\chi_{zz}$ [Sec.~\ref{app.sec:sumrule}] is measured, per site and
averaged over the two-site unit cell, by the estimator
\begin{equation}
\chi_{zz}=\frac12\left[1-\langle\sigma^z\rangle^2\right]
+\sum_{r=1}^{R}\overline{\langle\sigma^z_i\sigma^z_{i+r}\rangle_c},
\qquad R=100\text{--}300,
\end{equation}
where $\overline{\langle \dots \rangle}$ denotes the unit-cell average. The neglected tail is negligible at all quoted
points, and the result is independent of the bond dimension.

\subsection{$\gamma\to0$ extrapolation}
\begin{figure*}[t]
\centering
\includegraphics[width=0.95\textwidth]{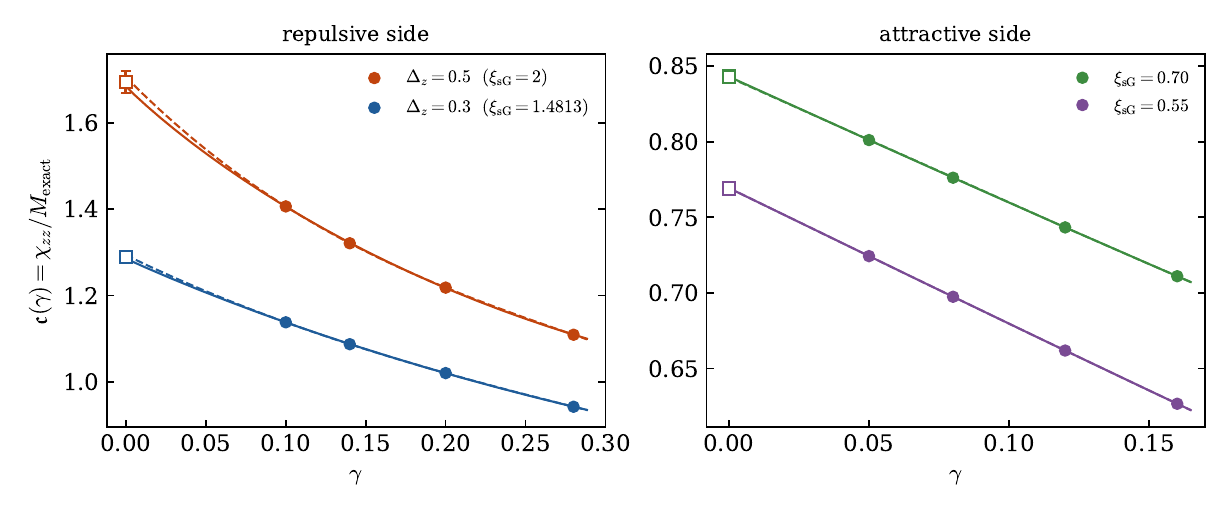}
\caption{$\gamma\to0$ extrapolations underlying Table~\ref{app.tab:grand}: iDMRG data (filled points),
cubic (dashed) and umklapp-corrected (solid) fits; open squares with error bars mark the adopted
$\gamma=0$ values. Left: repulsive side, where the extrapolation distance is largest; right:
attractive side.}
\label{app.fig.extrap}
\end{figure*}

$\mathfrak c(\gamma)=\chi_{zz}/M_{\rm exact}$ is extrapolated to $\gamma\to0$ from the four-point grids
below ($\gamma\in\{0.10,0.14,0.20,0.28\}$ repulsive, $\{0.05,0.08,0.12,0.16\}$ attractive) using an
ensemble of forms: two quadratics (first three points; all four), a cubic through all four, and the
constrained umklapp form $\mathfrak c(1+a\gamma+b\gamma^{s})$ with the fixed non-analytic
exponent $s=(4K-2)\pi/2\theta_0$ set by the Luttinger parameter $K=1/2\beta^2$ (no free exponent). The
umklapp form is the physically motivated one---it carries the leading lattice irrelevant operator absent
from the continuum---while the polynomials gauge the analytic corrections to scaling. The central value
is the mean of the cubic and umklapp extrapolants. The per-coupling fit coefficients and residuals are
provided with the code in the repository.
The data and fits are shown in Fig.~\ref{app.fig.extrap}. As a calibration, at $\Delta_z=0$ the same
procedure applied to the free-fermion ratio recovers unity. On the repulsive side the extrapolated limits
sit well above the smallest-$\gamma$ datum---and by a larger margin than on the attractive side---meaning that 
the extrapolation distance controls the spread, largest at
$\xi_{\rm sG}=2$. The susceptibility $\chi_{zz}$ is bond-dimension converged even at this hardest point, so the
dominant systematic is the residual model-dependence of the $\gamma\to0$ limit. Reaching smaller $\gamma$
on the repulsive side is limited by the growing correlation length: keeping the neglected correlator
tail small then requires a longer summation range $R$, the bond dimension being already converged.
The attractive anisotropies are the couplings implied by $\xi_{\rm sG}$,
$\Delta_z=\cos[\pi(1-\beta^2)]=-0.273663$ $(\xi_{\rm sG}=0.70)$ and $-0.440394$ $(\xi_{\rm sG}=0.55)$.
The complete finite-$\gamma$ data for all eight couplings of Table~\ref{app.tab:grand} follow, on the
grids and at the iDMRG parameters above:
\begin{widetext}
\begin{center}\footnotesize\setlength{\tabcolsep}{4pt}
\begin{tabular}{cc|cccc|cccc}
\hline\hline
& & \multicolumn{4}{c|}{$\chi_{zz}$} & \multicolumn{4}{c}{$\mathfrak c(\gamma)=\chi_{zz}/M_{\rm exact}$}\\
$\Delta_z$ & $\xi_{\rm sG}$ & 0.10 & 0.14 & 0.20 & 0.28 & 0.10 & 0.14 & 0.20 & 0.28\\\hline
$0.15$ & 1.212 & 0.069866 & 0.097432 & 0.136642 & 0.184766 & 1.0071 & 0.9672 & 0.9120 & 0.8449\\
$0.30$ & 1.481 & 0.052440 & 0.075902 & 0.110552 & 0.154624 & 1.1379 & 1.0872 & 1.0202 & 0.9422\\
$0.40$ & 1.710 & 0.042492 & 0.063318 & 0.094955 & 0.136268 & 1.2530 & 1.1895 & 1.1088 & 1.0185\\
$0.50$ & 2.000 & 0.033724 & 0.051977 & 0.080607 & 0.119101 & 1.4065 & 1.3211 & 1.2183 & 1.1091\\\hline
$\Delta_z$ & $\xi_{\rm sG}$ & 0.05 & 0.08 & 0.12 & 0.16 & 0.05 & 0.08 & 0.12 & 0.16\\\hline
$-0.15$ & 0.825 & 0.064798 & 0.096848 & 0.135364 & 0.170112 & 0.8636 & 0.8385 & 0.8055 & 0.7734\\
$-0.2737$ & 0.700 & 0.082813 & 0.120106 & 0.163532 & 0.201713 & 0.8011 & 0.7762 & 0.7433 & 0.7110\\
$-0.35$ & 0.629 & 0.096277 & 0.137168 & 0.183854 & 0.224232 & 0.7653 & 0.7398 & 0.7061 & 0.6729\\
$-0.4404$ & 0.550 & 0.115324 & 0.160919 & 0.211724 & 0.254770 & 0.7243 & 0.6974 & 0.6618 & 0.6266\\\hline\hline
\end{tabular}\end{center}
\end{widetext}
The masses $M(\gamma,\Delta_z)$ are those of
Sec.~\ref{app.sec:deriv}; each $\mathfrak c(\gamma)$ column is the input to the
$\gamma\to0$ fits.

\begin{table*}[t]
\centering
\begin{tabular}{c c c c c c c}
\hline\hline
$\xi_{\rm sG}$ & $\beta^2$ & $\mathfrak c_{\rm meas}$ & $\mathfrak c_{2\rm k}{+}\mathfrak c_{B_1}$ &
$10^3\,r_{-1}$ & $\mathfrak c_{\rm low}$ & $\mathfrak c_{\rm up}$ \\\hline
0.55 & 0.355 & $0.769$ & $0.75883$ & $0.71$ & $0.76132$ & $0.76981$\\
0.6291 & 0.386 & $0.808$ & $0.79886$ & $0.47$ & $0.80060$ & $0.80929$\\
0.70 & 0.412 & $0.843$ & $0.83562$ & $0.30$ & $0.83677$ & $0.84501$\\
0.8251 & 0.452 & $0.906$ & $0.90247$ & $0.099$ & $0.90285$ & $0.90887$\\
1.00 & 0.500 & $1$ & $1$ & $0$ & $1$ & $1$\\
1.2120 & 0.548 & $1.119$ & $1.12424$ & $0.12$ & $1.12472$ & --\\
1.4813 & 0.597 & $1.289$ & $1.29092$ & $0.54$ & $1.29309$ & --\\
1.7100 & 0.631 & $1.454$ & $1.43973$ & $1.05$ & $1.44394$ & --\\
2.00 & 0.667 & $1.695$ & $1.63742$ & $1.81$ & $1.64468$ & --\\
\hline\hline
\end{tabular}
\caption{iDMRG measurements of $\mathfrak c$ at eight couplings, using exact masses, against the lower bound and the corridor. Here $r_{-1}$ is the tail moment of Eq.~\eqref{app.eq.Tm1exact} minus the two-kink$+B_1$ contribution;
$\mathfrak c_{\rm low}=\mathfrak c_{2\rm k}+\mathfrak c_{B_1}+E_{\rm th}r_{-1}$ bounds $\mathfrak c$ from below at every coupling and $\mathfrak c_{\rm up}=\mathfrak c_{2\rm k}+\mathfrak c_{B_1}+ \sqrt{r_{+1}r_{-1}}$ bounds it from above for $\xi_{\rm sG}<1$ [Eq.~\eqref{app.eq.corridor2}]. The bound values are evaluated with two independent quadratures in close agreement.}
\label{app.tab:grand}
\end{table*}

\subsection{Direct evaluation of the asymmetry}
For right-canonical tensors $B$ and left Schmidt weights $\lambda$,
$\rho_A(\vec s,\vec t)=\sum_{ab}\lambda_a^2[\prod B]^{\vec s}_{ab}[\prod\bar B]^{\vec t}_{ab}$ for a
bulk interval, and
\[
\Tr[\rho_A e^{i\alpha Q}\rho_A e^{-i\alpha Q}]
=\sum_{a,c}\lambda_a^2\lambda_c^2\sum_{b,d}\big|E^{(\ell)}_\alpha[(ac),(bd)]\big|^2,
\]
with $E^{(\ell)}_\alpha=\mathcal T_\alpha^{\ell}$ the $\ell$-th power of the phase-dressed doubled
transfer operator of Sec.~\ref{app.sec.charged_moments}. One incremental pass per $\alpha$ yields all $\ell$,
and a uniform $\alpha$-grid of $N\ge2\ell+2$ circle points is achieved (Gaussian decay of the charge
distribution allows $N\sim40$--$64$ in practice). Validated against explicitly constructed interval
density matrices with charge projectors. Extracted constant at a representative interacting point:
$a+b/\ell$ fits of $c_1(\ell)=\Delta S^{(2)}-\frac12\log \ell$ over a range of fitting windows give a
consistent value $c_1\approx-0.1834$, the mild drift across windows reflecting the
$\mathcal O((M\ell)^{-1})$ tail at the accessible $M\ell$. We repeat the extraction at a second
interacting point on the attractive side, $(\Delta_z,\gamma)=(-0.3,0.2)$, where the directly measured
$\chi_{zz}\approx0.244$: the same fit gives $c_1\approx0.211$, matching the master-formula value
$\frac12\log(2\pi\chi_{zz})\approx0.213$ and inverting to $\chi_{zz}=e^{2c_1}/2\pi\approx0.243$, in
agreement with the direct measurement. The $n=2$ constant is thereby confirmed independently on both
sides of the free-fermion point.

\begin{figure*}[t]
\centering
\includegraphics[width=0.92\textwidth]{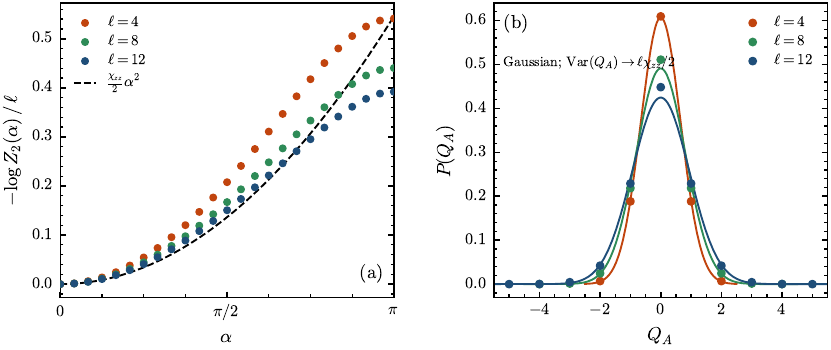}
\caption{Gaussian full counting statistics of the broken charge at the interacting point
$(\Delta_z,\gamma)=(0.3,0.2)$, from interval density matrices. (a)~Charged moments: the extensive
rate $-\log Z_2(\alpha)/\ell$ for $\ell=4,8,12$ collapses onto the parabola $(\chi_{zz}/2)\alpha^2$
(dashed) as $\ell$ grows---the Gaussian saddle that turns the width $V$ into
$\Delta S^{(2)}_A=\frac12\log(4\pi V)$, with $Z_2(\alpha)=1-V\alpha^2+\dots$ and $V=\ell\chi_{zz}/2$ to
leading order. (b)~The subsystem-charge distribution $P(Q_A)=\Tr[\Pi_{Q_A}\rho_A]$ broadens with $\ell$;
solid curves are Gaussians of the measured variance, which approaches $\ell\chi_{zz}/2$. Both panels are
governed by the single susceptibility $\chi_{zz}$ that fixes the entanglement asymmetry.}
\label{app.fig.fcs}
\end{figure*}
The same construction exposes the Gaussian structure that underlies the master formula directly:
Fig.~\ref{app.fig.fcs} shows the charged moments $Z_2(\alpha)$ collapsing onto the extensive rate
$(\chi_{zz}/2)\alpha^2$ and the charge distribution $P(Q_A)$ approaching a Gaussian of variance
$\ell\chi_{zz}/2$, both fixed by the susceptibility.

\bibliographystyle{apsrev4-2}
\bibliography{refs}

@article{AMC23,
  author = {Ares, Filiberto and Murciano, Sara and Calabrese, Pasquale},
  title = {Entanglement asymmetry as a probe of symmetry breaking},
  journal = {Nat. Commun.}, volume = {14}, pages = {2036}, year = {2023},
  eprint = {2207.14693}, archivePrefix = {arXiv}}

@article{RKACMB24,
  author = {Rylands, Colin and Klobas, Katja and Ares, Filiberto and Calabrese, Pasquale and Murciano, Sara and Bertini, Bruno},
  title = {Microscopic origin of the quantum {M}pemba effect in integrable systems},
  journal = {Phys. Rev. Lett.}, volume = {133}, pages = {010401}, year = {2024},
  eprint = {2310.04419}, archivePrefix = {arXiv}}

@article{Joshi24,
  author = {Joshi, Lata Kh. and others},
  title = {Observing the quantum {M}pemba effect in quantum simulations},
  journal = {Phys. Rev. Lett.}, volume = {133}, pages = {010402}, year = {2024},
  eprint = {2401.04270}, archivePrefix = {arXiv}}

@article{Khor24,
  author = {Khor, Brayden J. J. and K{\"u}rk{\c{c}}{\"u}oglu, Do{\u{g}}a M. and Hobbs, T. J. and Perdue, Gabriel N. and Klich, Israel},
  title = {Confinement and kink entanglement asymmetry on a quantum {I}sing chain},
  journal = {Quantum}, volume = {8}, pages = {1462}, year = {2024},
  eprint = {2312.08601}, archivePrefix = {arXiv}}

@article{MAKC23,
  author = {Murciano, Sara and Ares, Filiberto and Klich, Israel and Calabrese, Pasquale},
  title = {Entanglement asymmetry and quantum {M}pemba effect in the {XY} spin chain},
  journal = {J. Stat. Mech.}, pages = {013103}, year = {2024},
  eprint = {2310.07513}, archivePrefix = {arXiv}}

@article{FAC23,
  author = {Ferro, Federica and Ares, Filiberto and Calabrese, Pasquale},
  title = {Non-equilibrium entanglement asymmetry for discrete groups: the example of the {XY} spin chain},
  journal = {J. Stat. Mech.}, pages = {023101}, year = {2024},
  eprint = {2307.06902}, archivePrefix = {arXiv}}

@article{Lastres24,
  author = {Lastres, Marta and Murciano, Sara and Ares, Filiberto and Calabrese, Pasquale},
  title = {Entanglement asymmetry in the critical {XXZ} spin chain},
  journal = {J. Stat. Mech.}, pages = {013107}, year = {2025},
  eprint = {2407.06427}, archivePrefix = {arXiv}}

@article{CM23,
  author = {Capizzi, Luca and Mazzoni, Michele},
  title = {Entanglement asymmetry in the ordered phase of many-body systems: the {I}sing field theory},
  journal = {J. High Energy Phys.}, volume = {12}, number = {2023}, pages = {144}, year = {2023},
  eprint = {2307.12127}, archivePrefix = {arXiv}}

@article{HCCA22,
  author = {Horv{\'a}th, D{\'a}vid X. and Calabrese, Pasquale and Castro-Alvaredo, Olalla A.},
  title = {Branch point twist field form factors in the sine-{G}ordon model {II}: Composite twist fields and symmetry resolved entanglement},
  journal = {SciPost Phys.}, volume = {12}, pages = {088}, year = {2022},
  eprint = {2105.13982}, archivePrefix = {arXiv}}

@article{Luk97,
  author = {Lukyanov, Sergei},
  title = {Form factors of exponential fields in the sine-{G}ordon model},
  journal = {Mod. Phys. Lett. A}, volume = {12}, pages = {2543}, year = {1997},
  eprint = {hep-th/9703190}, archivePrefix = {arXiv}}

@article{FT11,
  author = {Feh{\'e}r, G{\'a}bor and Tak{\'a}cs, G{\'a}bor},
  title = {Sine-{G}ordon form factors in finite volume},
  journal = {Nucl. Phys. B}, volume = {852}, pages = {441}, year = {2011},
  eprint = {1106.1901}, archivePrefix = {arXiv}}

@article{LZ97,
  author = {Lukyanov, Sergei and Zamolodchikov, Alexander},
  title = {Exact expectation values of local fields in the quantum sine-{G}ordon model},
  journal = {Nucl. Phys. B}, volume = {493}, pages = {571}, year = {1997},
  eprint = {hep-th/9611238}, archivePrefix = {arXiv}}

@article{Zam95,
  author = {Zamolodchikov, Al. B.},
  title = {Mass scale in the sine-{G}ordon model and its reductions},
  journal = {Int. J. Mod. Phys. A}, volume = {10}, pages = {1125}, year = {1995}}

@article{JKM73,
  author = {Johnson, J. D. and Krinsky, S. and McCoy, B. M.},
  title = {Vertical-arrow correlation length in the eight-vertex model and the low-lying excitations of the {X}-{Y}-{Z} {H}amiltonian},
  journal = {Phys. Rev. A}, volume = {8}, pages = {2526}, year = {1973}}

@book{Baxter,
  author = {Baxter, Rodney J.},
  title = {Exactly Solved Models in Statistical Mechanics},
  publisher = {Academic Press}, address = {London}, year = {1982}}

@article{EEFR12,
  author = {Ercolessi, Elisa and Evangelisti, Stefano and Franchini, Fabio and Ravanini, Francesco},
  title = {Correlation length and unusual corrections to entanglement entropy},
  journal = {Phys. Rev. B}, volume = {85}, pages = {115428}, year = {2012},
  eprint = {1201.6367}, archivePrefix = {arXiv}}

@article{EEFR11,
  author = {Ercolessi, Elisa and Evangelisti, Stefano and Franchini, Fabio and Ravanini, Francesco},
  title = {Essential singularity in the {R}{\'e}nyi entanglement entropy of the one-dimensional {XYZ} spin-1/2 chain},
  journal = {Phys. Rev. B}, volume = {83}, pages = {012402}, year = {2011},
  eprint = {1008.3892}, archivePrefix = {arXiv}}

@article{EER10,
  author = {Ercolessi, Elisa and Evangelisti, Stefano and Ravanini, Francesco},
  title = {Exact entanglement entropy of the {XYZ} model and its sine-{G}ordon limit},
  journal = {Phys. Lett. A}, volume = {374}, pages = {2101}, year = {2010},
  eprint = {0905.4000}, archivePrefix = {arXiv}}

@article{tenpy,
  author = {Hauschild, Johannes and Pollmann, Frank},
  title = {Efficient numerical simulations with tensor networks: {T}ensor {N}etwork {P}ython ({TeNPy})},
  journal = {SciPost Phys. Lect. Notes}, pages = {5}, year = {2018},
  eprint = {1805.00055}, archivePrefix = {arXiv}}

@article{RVC24,
  author = {Rylands, Colin and Vernier, Eric and Calabrese, Pasquale},
  title = {Dynamical symmetry restoration in the {H}eisenberg spin chain},
  journal = {J. Stat. Mech.}, pages = {123102}, year = {2024},
  eprint = {2409.08735}, archivePrefix = {arXiv}}

@article{AresRev,
  author = {Ares, Filiberto and Calabrese, Pasquale and Murciano, Sara},
  title = {The quantum {M}pemba effects},
  journal = {Nat. Rev. Phys.}, volume = {7}, pages = {451}, year = {2025},
  eprint = {2502.08087}, archivePrefix = {arXiv}}

@article{Fossati24,
  author = {Fossati, Michele and Ares, Filiberto and Dubail, J{\'e}r{\^o}me and Calabrese, Pasquale},
  title = {Entanglement asymmetry in {CFT} and its relation to non-topological defects},
  journal = {J. High Energy Phys.}, volume = {05}, number = {2024}, pages = {059}, year = {2024},
  eprint = {2402.03446}, archivePrefix = {arXiv}}

@article{DdV95,
  author = {Destri, Claudio and de Vega, Hector J.},
  title = {Unified approach to thermodynamic {B}ethe {A}nsatz and finite size corrections for lattice models and field theories},
  journal = {Nucl. Phys. B}, volume = {438}, pages = {413}, year = {1995},
  eprint = {hep-th/9407117}, archivePrefix = {arXiv}}

@article{KW78,
  author = {Karowski, Michael and Weisz, Peter},
  title = {Exact form factors in (1+1)-dimensional field theoretic models with soliton behaviour},
  journal = {Nucl. Phys. B}, volume = {139}, pages = {455}, year = {1978}}

@article{Seiberg90,
  author = {Seiberg, Nathan},
  title = {Notes on quantum {L}iouville theory and quantum gravity},
  journal = {Prog. Theor. Phys. Suppl.}, volume = {102}, pages = {319}, year = {1990}}

@article{CV23,
  author = {Capizzi, Luca and Vitale, Vittorio},
  title = {A universal formula for the entanglement asymmetry of matrix product states},
  journal = {J. Phys. A}, volume = {57}, pages = {45LT01}, year = {2024},
  eprint = {2310.01962}, archivePrefix = {arXiv}}

\end{document}